\documentclass[12pt]{article}
\usepackage{amssymb}
\usepackage{amsmath}
\usepackage[ps,dvips,matrix,arrow,frame,import,curve,color]{xy}
\usepackage{subfigure, axodraw}
\usepackage{graphicx}
\makeatletter
\@addtoreset{equation}{section}
\makeatother

\def\be{\begin{equation}}
\def\ee{\end{equation}}

\def\ba{\begin{array}}
\def\ea{\end{array}}

\newcommand{\bea}{\begin{eqnarray}}
\newcommand{\eea}{\end{eqnarray}}
\def\N{$\cal N$}

\begin{document}
\hfill{}

\begin{flushright}
SU-ITP-10/17\\
DCPT-10/15
\end{flushright}

\vskip 2cm

\vspace{24pt}

\begin{center}
{ \LARGE {\bf    New \N=4 SYM Path Integral
 }}

\vspace{24pt}

{\large  {\bf  Chih-Hao Fu${}^\dag$ and   Renata Kallosh${}^\ddag$}}

    \vspace{15pt}

{${}^\ddag$Department of Physics, Stanford University, Stanford, CA 94305}

{${}^\dag$Department of Mathematical Sciences, University of Durham\\
South Road, Durham, DH1 3LE, U.K. }

\footnotetext{ E-mails:
${}^\dag$chih-hao.fu@durham.ac.uk, ${}^\ddag$kallosh@stanford.edu}

\vspace{10pt}

\vspace{24pt}

\end{center}

\begin{abstract}

Using Lorentz covariant spinor helicity formalism we reorganize the unitary scalar superfield light-cone path integral for the \N=4 supersymmetric Yang-Mills theory.  In new variables in the chiral Fourier superspace the quadratic and cubic parts of the classical action  have manifest Lorentz, kinematical and dynamical supersymmetry,  with the exception of terms which contribute only to the contact terms in the supergraphs with  propagators shrinking to a point. These terms have the same structure as    supergraphs with quartic light-cone vertices,  which
break  dynamical  supersymmetry.
We present evidence that  all complicated terms breaking dynamical supersymmetry have to  cancel and therefore can be omitted. 
It is  plausible that the new form of the path integral leads to a set of relatively simple unitarity based rules  with manifest \N=4  supersymmetry.

\end{abstract}
\newpage
\tableofcontents
\newpage
\section{Introduction}

The unitary light-cone superfield path integral  for \N=4 SYM is based on the light-cone superfield action \cite{Brink:1982pd}, \cite{Mandelstam:1982cb}, \cite{Brink:1982wv}, \cite{Belitsky:2004sc}.  
The light-cone superfield action correspond to the choice of the gauge $A_+=0$ for the vector field. It has a manifest kinematical supersymmetry, however, the dynamical and Lorentz symmetries are not manifest. In the Lorentz-covariant gauges there are 16 supersymmetries, $\bar q_{\dot \alpha }^A$ and $ q_{\alpha A}$, where $\alpha, \dot \alpha =1,2$ and $A, B=1, ..., 4$. These 16 supersymmetries are split in a Lorentz non-covariant way into 8+8. The first 8, $ \bar q_{\dot 2 }^A$ and $ q_{ 2 B}$,  are realized manifestly by introducing 8 Grassmann coordinates  in the light-cone superspace, $\theta^A$ and $\bar \theta_A$.  These are called kinematical supersymmetries. They are manifest since the action is given by an integral over 8 $\theta$, $\bar \theta$ of the Lagrangian which depends on 
 the light-cone  superfield $\Phi(x, \theta, \bar \theta)$ and its supercovarinat derivatives.
 The commutator of two kinematical supersymmetries is equal to $p_+$, i. e. $ \{\bar q_{\dot 2 }^A,  q_{2 B}\}= \delta^A{}_B \, p_+ $.  The light-cone superspace does not have Grassmann coordinates  associated with dynamical supersymmetry charges  $ \bar q_{\dot 1 }^A$ and $  q_{1 B}$. They are called dynamical since the commutator of two dynamical supersymmetries is the light-cone Hamiltonian, $ \{\bar q_{\dot 1 }^A,  q_{1 B}\}= \delta^A{}_B \, {p_\bot \bar p_\bot\over p_+} =\delta^A{}_B \, p_- $ for massless particles.

 The  Feynman rules for the light-cone superfields are complicated, not Lorentz covariant, and they were used mostly to prove the finiteness of the theory 
\cite{Mandelstam:1982cb}, \cite{Brink:1982wv} rather than for practical computations.

Here we will try to find a simpler approach to the light-cone path integral relating the action to the generalization of the Nair-type \cite{Nair:1988bq} off shell superfield developing the proposal  in \cite{Kallosh:2009db},\cite{Fu:2009cg}. The derivation of the new form of the unitary \N=4 path integral in momentum superspace requires few steps.

 Starting from from the light-cone superfield path integral based on the light-cone superfield actions \cite{Brink:1982pd}, \cite{Mandelstam:1982cb}, \cite{Belitsky:2004sc} the proposal  requires a) a change of the Grassmann variable $\pi ={\partial \over \partial \theta}$ into  a dimensionless one $\eta = \pi (\sqrt {p_+})^{-1}$  b) a Fourier transform from $(x, \theta)$ into a chiral Fourier superspace $(p, \eta)$ c) a rescaling of the original superfield $\Phi(x, \theta, \bar \theta)$ by a factor $p_+$. In this way  Lorentz non-covariant factors $p_+$, $p_\bot$, $\bar p_\bot$ of the cubic vertices
 are absorbed into the spinor helicity brackets.  The cubic vertices turn out to be given by the off shell generalization of the familiar 3-point MHV and   ${\overline{\text{MHV}}}$ amplitudes depending on Lorentz covariant angular and square helicity brackets\footnote{A particular off shell continuation of the vertices in the spinor helicity formalism was suggested for the computation of the  tree amplitudes in Yang-Mills theory in  \cite{Cachazo:2004kj}.}. In this way,  the quadratic and cubic part of the action in new variables becomes Lorentz covariant, up to a controllable part: ${\cal S}^2+ {\cal S}^3 = {\cal S}_{cov}^2+{\cal S}^3_{cov} + {\cal S}^3_{\Delta}+{\cal S}^3_{\bar \Delta}$. The quartic part of the action ${\cal S}^4$ remains complicated.

We split the Feynman rules into a part (i) where only the simple cubic vertices from ${\cal S}^3_{cov}=
\tilde{\mathcal{S}}^{3}_{\text{MHV}}+\tilde{\mathcal{S}}^{3}_{\overline{\text{MHV}}}$ are used, which have manifest kinematical and  dynamical supersymmetry and Lorentz symmetry, and the rest, part (ii). This part (ii) is complicated, it involves   cubic vertices from  ${\cal S}^3_{\Delta}$, ${\cal S}^3_{\bar \Delta}$ part of the action and the quartic ones, from ${\cal S}^4$. Each vertex  in part (ii)  can be shown to break dynamical supersymmetry.

Since the computation using only the  covariant cubic vertices from ${\cal S}^3_{cov}$ are relatively simple, the major problem is to find out  the role of the part (ii) in the computations of the on shell amplitudes. Are all terms in (ii), which individually break dynamical and Lorentz symmetry, canceling or do they combine into a non-trivial part of the on shell amplitudes? If they would cancel, it would mean that it is safe to perform the computations with rather simple Feynman rules in part (i) which have manifest supersymmetry and Lorentz symmetry. However, if the terms in (ii) combine into a covariant non-vanishing expressions, it would mean that the simple part of the new path integral is incomplete and one has to compute also the supergraphs with rather complicated vertices.

We will perform here a computation of the 4-point tree amplitude as a sample of the computation using the new path integral Feynman rules. We will find out that part (i) easily gives the correct answer in this particular computation.  This 4-point computation gives an evidence that the split of the Feynman rules into a simple covariant part and complicated non-covariant one may be valid also for tree amplitudes with more legs and  more loops. If the complicated supersymmetry breaking part drops from the result as it does in our example of the 4-point tree amplitude, it would mean that the new path integral may become an important tool for the maximally supersymmetric QFT. Much more computations will be required to check if the simple trend in the 4-point tree supergraph computations will remain valid in more complicated cases.

It would be interesting also to compare the new path integral rules with the unitarity cut method \cite{Bern:1994zx} which was used successfully in all most advanced computations of the higher loop diagrams in \N=4 SYM and \N=8 supergravity \cite{Bern:2009kd}.

\section{New form of the \N=4  supersymmetric path integral}

 In the Brink-Lindgren-Nilsson formalism \cite{Brink:1982pd} the light-cone action in the real superspace basis has terms which are quadratic, cubic and quartic in chiral and anti-chiral superfields, $\Phi(x, \theta,\bar \theta)$ and $\bar \Phi(x, \theta, \bar \theta)$, respectively.
\be
S[\Phi, \bar \Phi] = S^2 + S^3 + S^4
\label{action}\ee
where
\begin{equation}
S^2 + S^3= \int d^{4}x\, d^{4}\theta\, d^{4}\bar{\theta}\, \left [ \bar{\Phi}^a \frac{\Box}{2 \, \partial^{+2}}\Phi^a  -\frac{2}{3} g f^{abc} \left(\frac{1}{\partial_{+}}\bar \Phi^a  \Phi^b\bar \partial {\Phi}^c +\frac{1}{\partial_{+}}{\Phi}^a \bar \Phi^b {\partial}\bar \Phi^c \right)\right ]
\hspace{4cm}
\label{S2S3}\end{equation}
\begin{eqnarray}
S_{4} = -\frac{1}{2}g^{2}f^{abc}f^{ade}\int d^{4}x\, d^{4}\theta\, d^{4}\bar{\theta}\,[ \frac{1}{\partial_{+}}(\Phi^{b}\partial_{+}\Phi^{c})\frac{1}{\partial_{+}}(\bar{\Phi}^{d}\partial_{+}\bar{\Phi}^{e}) +{1\over 2}\Phi^{b}\bar{\Phi}^{c}\Phi^{d}\bar{\Phi}^{e}]
\label{S4}\end{eqnarray}
One can  see that Lorentz symmetry is broken in the corresponding Feynman rules. 
Due to the CPT invariance of the \N=4 supermultiplet the anti-chiral superfield is  related to the chiral one as follows
\be
\bar \Phi(x, \theta^A, \bar \theta_A)= -{1\over 4!} \partial_+ ^{-2} \epsilon ^{ABCD} D_A D_B D_C D_D \Phi(x, \theta^A, \bar \theta_A)
\label{constraint}\ee
Here $\Phi=\Phi^a t^a$ with $t^a$ being the generators  in the fundamental representation of the $SU(N)$ group.

We now define the following Fourier transforms of the light-cone superfields consistent with the constraint
({\ref{constraint}):
\begin{equation}
\Phi(x,\theta,\bar{\theta})=e^{\frac{1}{2}\bar{\theta}\cdot\theta\partial_{+}}\int\frac{d^{4}p}{(2\pi)^{4}}d^{4}\eta\, e^{ip\cdot x+\eta\frac{p_{+}}{\sqrt{p_{+}}}\theta}\,\left(\frac{-i}{p_{+}}\right)\phi(p,\eta)\label{eq:phi}\end{equation}

\begin{equation}
\bar{\Phi}(x,\theta,\bar{\theta})=e^{-\frac{1}{2}\bar{\theta}\cdot\theta\partial_{+}}\int\frac{d^{4}p}{(2\pi)^{4}}d^{4}\eta\, e^{ip\cdot x}\,\delta^{4}(\bar{\theta}\sqrt{p_{+}}-i\eta)\left(\frac{-i}{p_{+}}\right)\phi(p,\eta)\label{eq:phibar}\end{equation}
 The  Lie-algebra valued off-shell superfield $\phi(p, \eta)= \phi^a(p, \eta)  t^a$ depends only on physical degrees of freedom of \N=4 SYM theory:
\bea
 \phi  = \bar A(p) + \eta_A \psi ^A (p) + {1\over 2!} \eta_{A} \eta_B  \phi ^{AB}(p) + {1\over 3!} \epsilon ^{ABCD} \eta_{A}\eta_B \eta_C  \bar \psi_D (p)+ {1\over 4!} \epsilon ^{ABCD} \eta_{A}\eta_B \eta_C  \eta_D  A(p)
\label{PhiYM}\eea

When $p^2=2(p_+ p_- - p_\bot \bar p_\bot )= 0$ this superfield is well known and is used to describe the super-wavefunction of the physical state with the helicity +1. For the path integral where $ \phi (p, \eta)$ is the integration variable $p^2 \phi(p, \eta)\neq 0$.

 When the Fourier transform of the light-cone action (\ref{action}) is performed, one finds that the new form
 of ${\cal S}^2+{\cal S}^3$ depending on the Fourier superspace superfield $\phi(p, \eta)$ is unexpectedly simple, whereas the quartic ${\cal S}^4$ terms remains complicated.

 The new path integral for the generating functional of the on shell amplitudes is given by
 \be
 \exp {i W[\phi_{in}(z) ]}= \int d\phi~  \exp {i \left ({\cal S} [ \phi(z)] + tr \int d^8 z\,   \phi_{in}(z) \, p^2\,  \phi(-z)\right )} \ .
\label{pathintYM}\ee
where
$
p^2 \phi(p, \eta)\neq 0$ and $ p^2 \phi_{in}(p, \eta)=0
$
 and
$z=(p, \eta)$ is  the 4+4 momentum superspace.  The integration is defined as
$
 d^8 z \equiv  {d^4 p\over (2\pi)^4} d^4 \eta
$.
The action in (\ref{pathintYM}) has terms which are quadratic, cubic and quartic in superfields $\phi$
\be
S[\Phi, \bar \Phi] = {\cal S }[ \phi(z)] = {\cal S}^2 + {\cal S}^3 + {\cal S}^4
\ee
We derive the new form of the action ${\cal S }[ \phi(z)]$  in Appendices B and C where the details of  the Fourier transformation from $S[\Phi, \bar \Phi]$ to ${\cal S }[ \phi(z)]$ are given. Below we present the answer for ${\cal S }[ \phi(z)] $.


\subsection{New ${\cal S}^2$ and  ${\cal S}^3$}
The quadratic part of the new action is
\be
{\cal S}^2 = -{1\over 2} tr  \int d^8 z\,  \phi (z) \, p^2\,  \phi (-z)=-  tr \int {d^4 p\over (2\pi)^4} \bar A \, p^2 \, A + ...
\label{0S2}\ee
Here the terms in  ... are  kinetic terms for the self-dual scalars $\phi^{AB}$ and spinors $\psi^A, \bar \psi _A$.

The cubic terms of the new action are also nice and simple:
\begin{equation}
{\cal S}^3 = tr   \int  \prod_{i=1}^{i=3} \{d^8 z_i  {\phi }(z_{i})\}  \left (V_{123} + \overline V_{123}\right)
 \label{S3}\end{equation}
 where
 \be
V_{123} = C { (2\pi)^4\delta^{(4)}(\sum_i p_{i}) \delta^8 (\sum_i \lambda_i \eta_i) \over \left\langle 12\right\rangle \left\langle 23\right\rangle \left\langle 31\right\rangle }
 \label{V}\ee
 and
\begin{equation}
\overline V_{123}=C  { (2\pi)^4\delta^{(4)}(\sum_i p_{i}) \delta^4 \left(\left[12\right]\eta_{3}+\left[23\right]\eta_{1}+\left[31\right]\eta_{2}\right)\over \left[12\right]\left[23\right]\left[31\right]}
\label{barV}\end{equation}
Here $C={ g \over 3} c_1 c_2 c_3$ and $c_i=sgn (p_{i+})$ for each of the outgoing momenta.
Note that in the off-shell cubic action in
(\ref{S3})-(\ref{barV})  the fields $ \phi(z_i)$ are not on shell, each of the three $p_i^2$ is an integration variable.  For a general,  not light-like  vector we will use two types of  spinors, $\lambda$ and $\xi$:
\be
p_{\alpha \dot \alpha} =( \xi_\alpha \bar \xi_{\dot \alpha} + \lambda_\alpha \bar \lambda_{\dot \alpha})
\ee
In the context of the light-cone superfields we define them in the Appendix A. One can also try to use the prescription for the off shell vertices proposed in \cite{Cachazo:2004kj} where
$
\lambda_\alpha= p_{\alpha \dot \alpha} \eta^{\dot \alpha}
$ and $ \eta^{\dot \alpha}$ is an arbitrary spinor, but here we will first start with the light-cone type prescription in the Appendix A.

The 3-vertices in (\ref{S3})-(\ref{barV}) have a  dependence on the $\lambda (p)$, $\bar \lambda (p)$ spinors as one can see in eqs. (\ref{V}), (\ref{barV}). However, there is also a dependence  on $\xi (p)$, $\bar \xi (p)$ spinors in of the form
\be
\delta^4\sum_i (p_i)_{\alpha \dot \alpha}= \delta^4 \left (\sum_i ( \xi_\alpha \bar \xi_{\dot \alpha} + \lambda_\alpha \bar \lambda_{\dot \alpha})_i\right)
\label{cons}\ee
It is therefore important to define the spinors $\lambda (p)$, $\bar \lambda (p)$ which form the helicity brackets and enter in the Grassmannian delta-functions in (\ref{V}), (\ref{barV}) as well as the spinors $\xi (p)$, $\bar \xi (p)$ which are required for the momentum conservation $\delta$-function. We  present the explicit definition of these spinors in terms of the components of momentum vector appropriate to the light-cone gauge\footnote{Note, however, that when the Feynman supergraphs are computed and the answer for the generating functional of the on-shell amplitudes,  $ \exp {i W[\phi_{in}(z) ]}$ in eq. (\ref{pathintYM}),  is obtained, it depends on $\lambda (p)$, $\bar \lambda (p)$ spinors for each of the external particles, which are on shell. In this expression there is no need to take any particular choice of the $\lambda (p)$, $\bar \lambda (p)$ spinors since the answer for the on shell amplitudes is Lorentz covariant.
} in the Appendix A.  In particular, we find that for the off shell fields with $p^2\neq 0$ there is a non-vanishing component $1\dot 1$ component of the bilinears of the $\xi$-spinors
\be
 \xi_1 \bar \xi_ {\dot 1}  = \left({p^2  \over \sqrt 2 p_+}\right)
\ee
This term is  absent when the on-shell supersymmetric 3-point amplitudes are constructed at the complex momenta as suggested in
\cite{Brandhuber:2008pf}, \cite{Drummond:2008bq}, \cite{ArkaniHamed:2008gz}. The corresponding amplitudes have instead of (\ref{cons}) only the $\lambda$-dependent part, since each $p^2_i=0$,
\be
\left (\delta^4\sum_i (p_i)_{\alpha \dot \alpha}\right )_{p_i^2=0} = \delta^4 \left (\sum_i ( \lambda_\alpha \bar \lambda_{\dot \alpha})_i\right)
\label{cons1}\ee

Upon $\eta$-integration our cubic action (\ref{S3})-(\ref{barV}) will produce all twelve 3-point couplings for the vector, spinor and scalar fields corresponding to the light-cone gauge, see for example eq. (3.13) in the first paper in \cite{Brink:1982pd}.


\subsection{New ${\cal S}^4$} 

The quartic  term  in the light-cone superfield action ${ S}^4$ in eq. (\ref{S4}) is not Lorentz invariant and not supersymmetric under dynamical supersymmetry, it is supersymmetric only under the kinematical supersymmetry.
It is useful to introduce here the following notation:
\be
\psi_{ ij} \equiv    \bar \lambda_{\dot 2 j} \eta_i  - \bar \lambda_{\dot 2 i} \eta_j
\ee
and keep in mind that in the light-cone gauge $\bar \lambda_{\dot 2}=2^{1/4} \sqrt p_+$.
 The 4-point interaction in the action consists of two parts
  \be
{\cal S} ^4 = {\cal S}_{4}^{1}+ {\cal S}_{4}^{2}=tr\int\left(\prod_{i=1}^{4}d^{8}z_{i}\phi_{i}\right)\, (V_{4}^{1}+ V_{4}^2),
\label{S4cal}\end{equation}
with 
\begin{eqnarray}
&&
V_{4}^{1}=-\frac{g^2}{16}\,\frac{(2\pi)^{4}\delta(\sum_{i}p_{i})\delta^{4}(\sum_{i}\lambda_{2 i} \eta_{i})}{p_{1+}p_{2+}p_{3+}p_{4+}}\,\frac{(p_{1+}-p_{2+})}{(p_{1+}+p_{2+})}\,\frac{(p_{3+}-p_{4+})}{(p_{3+}+p_{4+})}  
  \left(\delta^{4}(\psi_{12})+\delta^{4}(\psi_{34})\right)
   \nonumber \\
&& \hspace{1cm} +cycl.
\label{S41}\end{eqnarray}
and
\begin{eqnarray}
&&
V_{4}^{2}=-\frac{g^2}{32}\,\frac{(2\pi)^{4}\delta(\sum_{i}p_{i})\delta^{4}(\sum_{i}\lambda_{2 i} \eta_{i})}{p_{1+}p_{2+}p_{3+}p_{4+}}\left(\delta^{4}(\psi_{24})+\delta^{4}(\psi_{13})
-\delta^{4}(\psi_{14})-\delta^{4}(\psi_{23})+cycl. \right)
\label{S42}\end{eqnarray}

A direct inspection shows that it breaks dynamical supersymmetry as well as Lorentz symmetry.

\section{Kinematical, $q_2, \bar q_{\dot 2}$,  and dynamical, $q_1, \bar q_{\dot 1}$,  supersymmetry}

Spinor helicity formalism is often applied in case of all-outgoing particle conventions. This means that the
 4-momenta of some particles are negative since some particles are ingoing. To distinguish between particle and antiparticle spinors it has been suggested in \cite{Bern:2009xq} to consider an analytic continuation rule that the change of the momentum sign is realized together with the change of the holomorphic spinors sign, whereas the non-holomorphic spinors do not change the sign.
For our purpose here, starting from the light-cone superfield action,  it is convenient  to take an opposite version of the analytic continuation rule, namely
 \be
 p\rightarrow -p \ , \qquad \lambda(p) \rightarrow  \lambda(-p) ,  \qquad \bar \lambda(p) \rightarrow - \bar \lambda(p)
 \label{prescr}\ee
 Thus the non-holomorphic spinors change the sign, whereas the holomorphic spinors do not change the sign.
 In case of the light-cone gauge  we present the details of  such an analytic continuation in the Appendix A.

Consider the linear transformation of the fields in the action under 16 supersymmetries
\be
\delta \phi(p, \eta)= (\epsilon^{\alpha A} q_{A\alpha} + \bar \epsilon^ {\dot \alpha}_A \bar q_{\dot \alpha}^A )\phi(p, \eta)
=(\epsilon^A \eta_A + \bar \epsilon_{A}  {\partial \over \partial \eta_A} )\phi(p, \eta)\ .
\ee
Here
\be
q_{A\alpha} = \lambda_\alpha \eta_A\, , \qquad  \bar q^A_{\dot \alpha}= \bar \lambda_{\dot \alpha}{\partial \over \partial \eta_A}\, , \qquad \epsilon^A\equiv  \epsilon^{\alpha A} \lambda_\alpha \, , \qquad \bar \epsilon _A \equiv \bar \epsilon^{\dot \alpha}_A \bar \lambda _{\dot \alpha} \ .
\label{susy}\ee
In the light-cone formulation of  \cite{Brink:1982pd},
 the kinematical supersymmetry is $q_{A 2}$ and $  \bar q_{\dot 2}^A$ and the dynamical is $q_{A 1}$ and $  \bar q_{\dot 1}^A$ and
 \be
\{ \bar q^A_{\dot \alpha} , q_{B\alpha} \}=  \delta^A{}_B \, \lambda_\alpha \bar \lambda_{\dot \alpha }
 \ee
Consider now the supersymmetry variation of the product of $n$ chiral superfields
\be
\delta\prod _{i=1}^n \phi(p_i, \eta_i) = (\epsilon^{\alpha A} Q_{A\alpha} + \bar \epsilon^ {\dot \alpha}_A \overline Q_{\dot \alpha}^A ) \prod _i \phi(p_i, \eta_i)
\ee
 where
 \be
 Q_{A\alpha}\equiv \sum_{i=1}^{n} \lambda_{i \alpha}  \eta_{i A} \, ,\qquad
\overline Q_{\dot \alpha}^A \equiv \sum_{i=1}^n \bar \lambda_{i \dot \alpha}{\partial \over \partial \eta_{Ai}} \  ,
\ee

We will find below that ${\cal S}^2$ is invariant under all 16 supersymmetries but the cubic action ${\cal S}^3$ is invariant only under 8 kinematical supersymmetries.

The cubic  action consists of two parts ${\cal S}^3_{ {MHV}}$ and ${\cal S}^3_{\overline {MHV}}$. We will find that
${\cal S}^3_{ {MHV}}$ is invariant under 8+4 supersymmetries, $ Q_{A\alpha}$ and $\overline Q_{A \dot 2}$, however,  the remaining   4 dynamical supersymmetries $\overline Q_{A \dot 1}$ are broken off shell. For  ${\cal S}^3_{\overline {MHV}}$ the opposite is true, namely, it is invariant under 8+4 supersymmetries, $ \bar Q_{A \dot \alpha}$ and $Q_{A  2}$ but the 4 dynamical supersymmetries $Q_{A  1}$ are broken off shell. The terms which break the dynamical supersymmery $\overline Q_{A \dot 1}$ and $Q_{A  1}$ have very distinctive features which will be derived below and which allow to relate them to the contribution from the 4-vertices from ${\cal S}^4$.

We will show that ${\cal S}^4$ is invariant under the kinematical supersymmetry and breaks the dynamical one.
Here it is useful to remind that the on shell supersymmetric 4-point amplitude has the following dependence on $\eta$'s.
\be
\delta^4 (Q_2) \delta^4 (Q_1)= \delta^8 (Q_\alpha)
\ee
Therefore under 4 +4  supersymmetries, $Q_1$  and $Q_2$, it is manifestly supersymmetric. The action of the remaining   4+4 supersymmetriers $\overline Q_{\dot \alpha}$ on $ \delta^8 (Q_\alpha)$ produces $P_{\dot \alpha \alpha}$ which vanishes due to the momentum conservation.

We have found, however, that the ${\cal S}^4$ vertex depends on $\eta$'s via various combinations of the following functions
\be
\delta^4 (Q_2) \delta^4(\psi_{ij})
\ee
where 
\be
\psi_{ij}\equiv \bar \lambda_{\dot 2}^i \eta_j - \bar \lambda_{\dot 2}^j \eta_i
\ee
We will show below that the  variation under dynamical supersymmetries, $Q_1$  and   $\overline Q_{\dot 1}$, does not vanish even with an account of momentum conservation for the 4 on shell particles.

In the covariant formulation of the on shell amplitudes their  supersymmetry properties  were studied in detail in \cite{Elvang:2008na}, \cite{Bern:2009xq}. We are using analogous methods here, however, in addition, we have to put particular attention to the off shell properties of the vertices we study.

 \subsection{Supersymmetry of ${\cal S}^2$}
It will be convenient to rewrite ${\cal S}^2$ as
\be
{\cal S}^2 = -{1\over 2} tr \int \prod _{i=1}^{2} \{d^8 z_i\,  \phi (z_i)\} (2\pi)^4 \delta^4( p_1+p_2) \delta^4( \eta_1+\eta_2) \, p_1^2
\label{S2}\ee
Since $p_1$ equals $-p_2$, it means that $\lambda(p_1) = \lambda(p_2)$ in our prescription (\ref{prescr}). This means that $Q_{A\alpha}= \lambda_\alpha (p_1) \eta_{1A} + \lambda_\alpha (p_2) \eta_{2A}= \lambda_\alpha (p_1) (\eta_{1A}+\eta_{2A})$. Since $Q_{A\alpha} \delta^4( \eta_1+\eta_2) \sim (\eta_{1A}+\eta_{2A}) \delta^4( \eta_1+\eta_2)=0$, the $S^2$ part of the action has a manifest $Q$ supersymmetry.

Under $\overline Q$ transformations  we find that $\left (\bar \lambda (p_1) {\partial \over \partial \eta_{A1}}  + \bar \lambda (p_2) {\partial \over \partial \eta_{A2}} \right)\delta^4 ( \eta_1+\eta_2)=\bar \lambda (p_1) + \bar \lambda (p_2) $. Since $p_1=-p_2$ in our prescription (\ref{prescr}) we get  $\bar \lambda (p_1) + \bar \lambda (p_2) =0$, which proves the remaining supersymmetry of ${\cal S}^2$. The quadratic part of the action has unbroken 16 supersymmetries despite the superfields in (\ref{S2}) are off shell.

\subsection{Off shell broken dynamical supersymmetry of ${\cal S}^3_{MHV}$}

We start with ${\cal S}^3_{MHV} $:
\begin{equation}
{\cal S}^3_{MHV} = C \int  tr \prod_{i=1}^{i=3}  \{d^8 z_i  {\phi}(z_{i})\}(2\pi)^4\delta^{(4)}(\sum_i p_{i})
 {\delta^8 (\sum_i \lambda_{\alpha i } \eta_i) \over \left\langle 12\right\rangle \left\langle 23\right\rangle \left\langle 31\right\rangle }
 \label{1S3}\end{equation}
The action of $Q_{A \alpha}= \lambda_\alpha (p_1) \eta_{1A} + \lambda_\alpha (p_2) \eta_{2A}+ \lambda_\alpha(p_3) \eta_{3A}$  on $ \delta^8 (\sum_i \lambda_i \eta_i)$ gives zero. However, the action of
 $\overline Q=\sum_{i=1}^3 \bar \lambda_{i \dot \alpha}{\partial \over \partial \eta_{Ai}}$ is more complicated. 
 \be
 \overline Q_{\dot \alpha}^A   (\sum_{i=1}^3 \lambda_{\alpha i} \eta_{i B})= \sum_{i=1}^3   \bar{\lambda}_{\dot \alpha} \lambda_\alpha \, \delta^A{}_B
  \ee
If we would have a conservation of momenta in the form $\sum_{i=1}^3 \bar \lambda_{\dot \alpha i}  \lambda_{\alpha i} =0$ as in the case of the on shell superfields with $p_i^2=0$, we would have an unbroken supersymmetry. However, for the off-shell case we find that the $\overline Q_ {\dot 1}$ component of the dynamical supersymmetry variation does not vanish since according to off shell momentum conservation  $\delta^{(4)}(\sum_i p_{i}) $
\be
\sum_{i=1}^3 \bar \lambda_{\dot \alpha} \lambda_{\alpha i} = - \sum_{i=1}^3 \bar \xi_{\dot \alpha} \xi_{\alpha i} \, ,
\qquad
 \sum_{i=1}^3 \bar \xi_{\dot 1} \xi_{1 i}=  \sum_{i=1}^3 {p_i^2\over \sqrt{2} \, p_{+i}}
\ee

Thus the action is not invariant under  4 dynamical supersymmetries $ \overline Q_{\dot 1}^A$.
It is interesting that only if  $\sum_{i=1}^3 {p_i^2\over p_+}$ vanishes, the dynamical  supersymmetry is unbroken. The deviation from supersymmetry always include terms with ${p^2_i\over p_{+i}} \phi(z_i)$.
When the vertex from ${\cal S}^3_{MHV}$ is inserted in any Feynman graph, the terms which break supersymmetry involve
\be
{p_i^2}  T\left (\phi(p_i, \eta_i) \phi(p_j, \eta_j) \right) \delta^4(p_i+ p_j) \sim {p_i^2} \,{1\over p_i^2} \sim 1
\label{shrink}\ee
In coordinate space this correspond to
\be
{\Box_x } T\left (\phi(x ) \phi(y) \right) \sim   \delta^4(x-y)
\label{shrink1} \ee
Thus the Lorentz covariant propagator shrinks to a point and the terms which break dynamical supersymmetry have a structure of the contact terms. We will see that analogous structures come from the 4-vertex insertion.


 \subsection{Off shell broken dynamical supersymmetry of $ {\cal S}^3_{\overline {MHV}}$}

Now we study ${\cal S}^3_{\overline {MHV}}$.
\begin{equation}
{\cal S}^3_{\overline {MHV}} =  C \int tr \prod_{i=1}^{i=3} \{d^8 z_i  {\phi}(z_{i})\}(2\pi)^4\delta^{(4)}(\sum_i p_{i})
{\delta^4 \left(\left[12\right]\eta_{3}+\left[23\right]\eta_{1}+\left[31\right]\eta_{2}\right)\over \left[12\right]\left[23\right]\left[31\right]}
 \label{2S3}\end{equation}
The action of $\epsilon^{ \dot \alpha}_A  \overline Q^A_{\dot \alpha} $ on $\delta^4 \left(\left[12\right]\eta_{3}+\left[23\right]\eta_{1}+\left[31\right]\eta_{2}\right)$ produces an expression which vanishes due to Schouten identity without the use of the momentum conservation:
\be
[\epsilon 1][23] + [\epsilon 2][31] + [\epsilon 3][12]=0
\ee
However, the action of the $Q_{A\alpha}$ supersymmetry is only partially symmetric. One can reorganize
it as follows 
\be
 \sum_{i=1}^3 \lambda_\alpha (p_i) \eta_{i } = {\lambda_\alpha (p_1) \over [23]} \left(\left[12\right]\eta_{3}+\left[23\right]\eta_{1}+\left[31\right]\eta_{2}\right) +  \sum_{i=1}^3   \lambda_\alpha \bar \lambda _{\dot \alpha} (p_i) \, {\bar \lambda^{\dot  \alpha} (p_3) \eta_2 -\bar  \lambda^{\dot \alpha} (p_2) \eta_3 \over [23]}
 \ee
 The first term is clearly annihilated by $\delta^4 \left(\left[12\right]\eta_{3}+\left[23\right]\eta_{1}+\left[31\right]\eta_{2}\right)$, however, the second term is proportional to  $ \sum_{i=1}^3  \xi_\alpha  \bar{ \xi} _{\dot \alpha} (p_i)$. Thus the $Q_{A 1}$ dynamical supersymmetry is broken since  $ \sum_{i=1}^3  \xi_1 \bar{\xi} _{\dot 1} (p_i) = \sum_{i=1}^3 {p^2_i\over \sqrt 2 \, p_{i+}}$.

 When the vertex from ${\cal S}^3_{\overline {MHV}} $ is inserted in any Feynman graph, the terms which break supersymmetry have the same structure as shown in eqs. (\ref{shrink}), (\ref{shrink1}): the Lorentz covariant propagator shrinks to a point.
 Thus, the dynamical supersymmetry of the cubic part of the action is broken as follows.
\be
\delta_{\bar \epsilon^{\dot 1}_A} {\cal S}^3_{ {MHV}} \neq 0 \, , \qquad  \qquad \delta_{ \epsilon^{ 1 A} } {\cal S}^3_{\overline {MHV}} \neq 0
\ee
In both cases the variation is proportional to $\sum_{i=1}^3 {p^2_i\over p_{i+}}$.

 \subsection{Broken dynamical supersymmetry of ${\cal S}^4$}

The ${\cal S}^4$ vertices depend on $\eta$'s via various combinations of the following functions
\be
V_4 \sim \delta^4 (Q_2) \delta^4(\psi_{ij}) f^{[ij]} (p) \delta^4 (P_{\dot \alpha \alpha} )
\ee
The fermionic $\eta$-dependent antisymmetric in the particle  position  function $\psi_{ij}$ has an interesting simple property under dynamical supersymmetry transformation
\be
\overline Q_{dyn} \, \psi_{ij} = [ij]
\ee
or, in more detail
\be
\overline Q_{\dot 1} ( \eta_i \bar \lambda_{\dot 2 j}- \eta_j \bar \lambda_{\dot 2 i})= \bar \lambda_{\dot 1 i} \bar \lambda_{\dot 2 j} -  \bar \lambda_{\dot 1 j} \bar \lambda_{\dot 2 i}
 = [ij]
\ee
 This is in a sharp contrast with the 4-point fermionic  $\eta$-dependent  function which is invariant on dynamical supersymmetry for the configuration of 4 particles satisfying the conservation of on shell momenta condition. The corresponding function is
 \be
A_4^{inv} \sim \delta^4 (Q_2) \delta^4(Q_1) \delta^4 (P_{\dot \alpha \alpha} ) = \delta^8 (Q_{\alpha}) \delta^4 (P_{\dot \alpha \alpha} )
 \ee
 It has the property
\be
Q_{\alpha} A_4^{inv}  =0 \, , \qquad \overline Q_{\dot \alpha} A_4^{inv} =0
\ee
Direct inspection shows that $V_4$ is invariant under the  action of kinematical supersymmetries
\be
Q_{2} V_4  =0 \, , \qquad \overline Q_{\dot 2} V_4 =0
\ee
However, for dynamical supersymmetries we find
\be
Q_{1} V_4  =Q_1 \left (\delta^4  (Q_2 ) \delta^4 ( \psi_{ij} )f^{[ij]}\delta^4 (P_{\dot \alpha \alpha} ) \right) \sim \left\langle mn\right\rangle \eta_m \eta_n  \delta^4 (\psi_{ij}) f^{[ij]} (p) \neq 0
\ee
and
\be
 \overline Q_{\dot 1} V_4 \sim [ij]  f^{[ij]} (p)\neq 0
\ee
Thus the reason why $V_4$ breaks both dynamical supersymmetries is because the $\eta$-dependence in the covariant amplitude
 \be
A_4^{cov} \sim   \delta^4(Q_1)
 \ee
is replaced by a different $\eta$-dependence
  \be
V_4 \sim    \delta^4(\psi_{ij}) f^{[ij]} (p)
 \ee

\section{New Feynman rules}

Consider the  3-vertices (\ref{V}) and (\ref{barV}). As explained above,  they break  dynamical supersymmetry since $\sum_i  \lambda_{i\alpha} \bar{\lambda}_{i\dot \alpha}\sim \sum_i {p_i^2 \over p_{+i}}\neq 0$. If in the process of the computation of the Feynman supergraphs the 3-vertices would be split into the value taken at  $\sum_i {p^2_i\over p^+_i}=0$ and the rest it would be
\be
V_{123}= (V_{123})_{\sum_i {p^2_i\over p^+_i}=0} + \Delta_{123} (p_i,\eta_i)
\ee
where
\be
 \Delta (p_i,\eta_i)\sim \sum_i {p^2_i\over p^+_i} X
\ee
and $X_i$ is non-singular in $p_i^2$ so that at $p_i^2=0$, $ \Delta_{123} (p_i,\eta_i)=0$.
Same for $V_{\overline {123}}$
\be
V_{\overline  123}= (V_{\overline  123})_{\sum_i {p^2_i\over p^+_i}=0} +\overline  \Delta_{123} (p_i,\eta_i)
\ee
where
\be
\overline  \Delta (p_i,\eta_i)\sim  \sum_i {p^2_i\over p^+_i} \overline  X
\ee
and $\overline  X $ is non-singular in $p_i^2$ so that at $p_i^2=0$, $ \overline  \Delta_{123} (p_i,\eta_i)=0$.

The total action now has been reorganized to the following form
\be
{\cal S}_{tot}= {\cal S}^2+ \tilde {\cal S}^3_{\text{MHV}}+ \tilde {\cal S}^3_{\overline{\text{MHV}}} + {\cal S}^3_{\Delta} + {\cal S}^3_{\overline \Delta} + {\cal S}^4
\ee
Here ${\cal S}^2$ is given in eq. (\ref{S2}), $\tilde {\cal S}^3_{\text{MHV}}$, $\tilde {\cal S}^3_{\overline{\text{MHV}}}$  are defined as follows 
\begin{equation}
\tilde {\cal S}^3_{\text{MHV}} =C \int tr  \prod_{i=1}^{i=3} \{d^8 z_i {\phi}(z_{i})\} { \delta^{(4)}(\sum_i \lambda_{i}^{\alpha} \bar{\lambda}_{i}^{\dot \alpha})\,\delta^8 (\sum_i \lambda_i \eta_i) \over \left\langle 12\right\rangle \left\langle 23\right\rangle \left\langle 31\right\rangle }
 \label{eq:S3tilde}\end{equation}

 \begin{equation}
\tilde {\cal S}^3_{\overline{\text{MHV}}} =  C  \int tr \prod_{i=1}^{i=3} \{d^8 z_i {\phi}(z_{i})\}{  \delta^{(4)}(\sum_i \lambda_{i}^{\alpha} \bar{\lambda}_{i}^{\dot \alpha}) \delta^4 \left(\left[12\right]\eta_{3a}+\left[23\right]\eta_{1a}+\left[31\right]\eta_{2a}\right)\over \left[12\right]\left[23\right]\left[31\right]} \label{eq:barS3}\end{equation}
The terms in the action ${\cal S}^3_{\Delta} + {\cal S}^3_{\overline \Delta}$ are cubic in superfields and have at least one $p^2$ in each vertex. These are terms with $\xi_{1} \xi_{\dot 1}$ which we ignored in $\tilde {\cal S}^3_{\text{MHV}}$, $\tilde {\cal S}^3_{\overline{\text{MHV}}}$, they are proportional to ${p^2\over p_+}$. If the relevant leg is an external one with the on shell particle, these terms vanish. However, if the relevant field is an off shell field inside the graph, there is an effect explained in eqs. (\ref{shrink}), (\ref{shrink1}): the covariant propagator shrinks to a point. Analogous terms come from the ${\cal S}^4$ vertex as both are the so-called contact terms with at least 4 lines in a single vertex, apart from possible ${1\over p_+}$ singularities.

{\it We will  assume here  that the contribution from all supersymmetric terms gives the correct answer whereas the contribution from all supersymmetry and Lorentz breaking terms cancels.}

 If the assumption is correct it would indicate that one should only use the quadratic and supersymmetric parts of the cubic action in the computation. The complicated quartic action is designed to remove from the answer the leftovers from the non-supersymmetric parts of cubic vertices. So, if we neglect both, the rules are simple.

\section{Computation of the 4-point tree supergraph amplitude}
To compute the 4-point connected on shell amplitude given by the terms  quartic in $\phi_{in}$ in the generating function $W(\phi_{in})$ we have to consider the tree supergraphs using either one MHV and one ${\overline{\text{MHV}}}$, or both MHV, or both ${\overline{\text{MHV}}}$ 3-vertices as well as a quartic vertex.
Here we should keep in mind that all superghraphs for the  4-point tree amplitude produce  a complete tree level amplitude $\mathbb{A}_4^{\rm tree} (1,2,3,4)$. This amplitude can be decomposed as follows
\be
\mathbb{A}_4^{\rm tree} (1,2,3,4)= g^4 \sum_{{\cal P} (2,3,4)} tr[t^{a_1}  t^{a_2} t^{a_3} t^{a_4} ] A_4^{\rm tree} (1,2,3,4)
\label{total}\ee
The  generators  of the gauge group $t^{a_i}$  encode the color of each external leg 1,2,3,4 with color group indices $a_i$. The sum runs over all noncyclic permutations of legs, which is equivalent to all permutations keeping one leg fixed (here leg 1). In case of interest we have 6 permutations, namely
\be
(1234), \quad (1243), \quad (1324), \quad (1342), \quad (1423), \quad (1432)
\label{6}\ee
The first 2 cases can have only poles in $s_{12}=s_{34}$ the next 2  cases can have only poles in $s_{13}=s_{24}$ and the last 2 cases can have only poles in $s_{14}=s_{32}$. In each of these six tree supergraphs we would have a factor of ${1\over 2}$ if we would use the total vertex, MHV$+{\overline{\text{MHV}}}$. Equivalently, we may skip ${1\over 2}$ and consider graphs with the leg 1 only in the vertex MHV and the other vertex is ${\overline{\text{MHV}}}$ vertex. In such case we have 4 supergraphs for the cases with poles in $s_{12}=s_{34}$ and $s_{14}=s_{23}$. In each of these cases the nearest neighbors are 1 and 4 or 1 and 2  but not 1 and 3.  These 4 graphs will give contribution to each of the 4 partial amplitudes. If we want to have just one of them we have to compute only one graph, for example with  MHV vertex with 1, 4 , P and the  ${\overline{\text{MHV}}}$ vertex with 2, 3, P shown  in Fig. 1. The second has 2 and 3 in the opposite order. The third one has 1, 2, P MHV and 3, 4, P ${\overline{\text{MHV}}}$ vertex. The fourth one has the same as the previous one but  with the opposite order for 3, 4.

\subsection{One MHV and one  ${\overline{\text{MHV}}}$ vertices }

For the color ordered partial 4-point tree amplitude, as explained above,  we have to compute the supergraph shown  in Fig. 1, where there is an MHV vertex with 1, 4, P and the  ${\overline{\text{MHV}}}$ vertex with 2, 3, P.  In addition we have the corresponding part of the contact term, presented in Fig. 2.
Other supergraphs, as explained above, will contribute to other terms in eq. (\ref{total}).
\begin{figure}[!h]
     \centering
     \subfigure{
     \includegraphics[height=3.7cm]{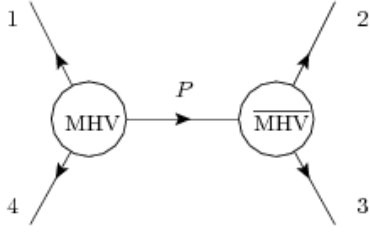} 
}
  \caption{The expression for this graph is presented in eq. (\ref{BHT1})}
  \label{fig1}
  \end{figure}

 \begin{figure}[!h]
     \centering
     \subfigure{
          \includegraphics[height=3.1cm]{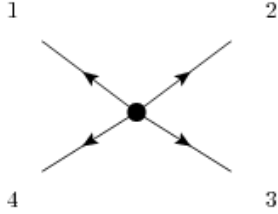} 
 }
  \caption{A contact term from the ${\cal S}^4$ vertex}
  \label{fig2}
  \end{figure}

The computation done in sec. (3.1) in \cite{Brandhuber:2008pf} based on the supersymmetric recursion relation is very close to the one we perform here. The difference is that our vertices are off shell and we split the expression into two parts, with and without $p_i^2$ terms. And in the Feynman integral we have to take into account more graphs, a priory. A special choice of the  shifts in super-momenta leads to significant reduction of the amount of supergraphs, only the superghraph in Fig. \ref{fig1} has to be computed.  It is also interesting that this single supergraph based on recursion relation gives the correct cyclic symmetric answer. Therefore in practical terms when there are more supergraphs in the path integral, for the case at hand, they give the same answer.

For the supergraph with an MHV vertex with 1, 4 , P and the  ${\overline{\text{MHV}}}$ vertex with 2, 3, P  in Fig. 1 we have to integrate over $P$ and $\eta_P$  the following expression
\be
{\delta^4(p_1+p_4+P) \delta^8 (\lambda^1\eta_1+ \lambda^4 \eta_4 + \lambda^P \eta_P ) \delta^4(-P+p_2+p_3)  \delta^4( \eta_P [23] + \eta_2[3P] +\eta_3 [P2])\over P^2 \left\langle 41\right\rangle [23] \left\langle 1P\right\rangle \left\langle P4\right\rangle [P2] [3P] }
\label{BHT1}\ee
Here the angular and square brackets are defined in eqs. (\ref{lambda}), (\ref{eq:spinors}) and do not require the on shell conditions since $p_-$ does not enter the definition of the brackets and does not have to be equal to ${p_\bot \, \bar p_{\bot}\over  p_+}$ for each particle.

Note that for the on shell $p_1, p_2, p_3, p_4$ we may use the momentum conservation  in the form  $\delta^4(( \lambda \bar\lambda)^2+( \lambda \bar\lambda)^3-( \lambda \bar \lambda)^P
-( \xi \bar \xi)^P)$.
The answer can be presented, as suggested in  the previous section, by splitting it into

 \noindent (i) part, where we replace  $\delta^4(p_2+p_3-P)$ by $\delta^4(( \lambda \bar\lambda)_2+( \lambda \bar\lambda)_3-( \lambda \bar\lambda)_P)$, corresponding to using only ${\cal S}^2+ \tilde {\cal S}^3_{\text{MHV}}+ \tilde {\cal S}^3_{\overline{\text{MHV}}} $, and

\noindent (ii) part,   which has a factor $(\xi \bar \xi_P)_{1\dot 1}= {P^2\over \sqrt 2 \, P_+} $
corresponding to using also  ${\cal S}_{\Delta} + {\cal S}_{\overline \Delta} + {\cal S}^4 $.


\subsubsection{Using only $\tilde{\mathcal{S}}^{3}_{\text{MHV}}$ and $\tilde{\mathcal{S}}^{3}_{\overline{\text{MHV}}}$ vertices}
Here we use only the part of the cubic vertices which is supersymmetric and is defined in eqs. (\ref{eq:S3tilde}), (\ref{eq:barS3}),
\begin{figure}[!h]
     \centering
     \subfigure{
\includegraphics[height=3.7cm]{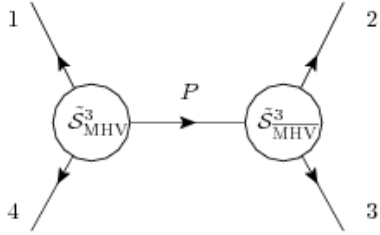}
 }
  \caption{The computation of this graphs leads to an answer in eq. (\ref{W4}). }
  \label{fig3}
  \end{figure}
We now use the fact that  in eq. (\ref{BHT1})
\be
\eta_P=  {\eta_2[P3] -\eta_3 [P2]\over [23]}
\ee
We insert this expression into $ \delta^8 (\lambda^1\eta_1+ \lambda^4 \eta_4 + \lambda^P \eta_P)$
and find
\be
 \delta^8 \left (\sum _{i=1}^4 \lambda^i \eta_i + {\eta_2 ( \lambda^P [P3] -\lambda^2 [23]) - \eta_3 ( \lambda^P  [P2] - \lambda^3 [32])  \over [23]} \right)
\ee
where we also added and subtracted $\lambda^2 \eta_2+ \lambda^3 \eta_3$. We rearrange the argument of the fermionic $\delta$-function as follows
\be
 \delta^8 \left (\sum _{i=1}^4 \lambda^i \eta_i + {\eta_2 ( \lambda^P \bar \lambda^P -\lambda^2 \bar \lambda^2 -  \lambda^3 \bar \lambda^3) \bar \lambda^3 - \eta_3 ( \lambda^P \bar \lambda^P -\lambda^2 \bar \lambda^2 -  \lambda^3 \bar \lambda^3) \bar \lambda^2 \over [23]} \right)
\label{fermD}\ee
If according to the prescription (i) and eqs. (\ref{eq:S3tilde}), (\ref{eq:barS3})  we use 
$( \lambda \bar\lambda)^2+( \lambda \bar\lambda)^3-( \lambda \bar\lambda)^P=0$, the fermionic $\delta$-function (\ref{fermD}) becomes
\be
 \delta^8 \left (\sum _{i=1}^4 \lambda^i \eta_i  \right)
\ee
Using the same prescription  $( \lambda \bar\lambda)^2+( \lambda \bar\lambda)^3-( \lambda \bar\lambda)^P=0$ we now simplify the remaining terms in (\ref{BHT1}), namely perform the replacements
\be
\left\langle 1P\right\rangle [P3]= \left\langle 12\right\rangle [23] \, , \qquad \left\langle 4P\right\rangle [P2]= \left\langle 43\right\rangle [32] \label{1}\ee
and integrate over $P$ and $\eta_P$.  Note also  that
\be
 {[23]^4\over  \left\langle 23\right\rangle [23] \left\langle 41\right\rangle [23] \left\langle 12\right\rangle \left\langle 34\right\rangle [23] [23] } = {1\over \left\langle 12\right\rangle \left\langle 23\right\rangle \left\langle 34\right\rangle \left\langle 41\right\rangle }
\label{BHTanswer}\ee
This reproduces the correct 4-point on shell amplitude, the quartic in free fields $\phi_{in}$ tree answer
\be
W^4(\phi_{\rm in})= g^2 tr \int \prod_{i=1}^4 \{d^8 z_i \phi_{\rm in}(z_i) \delta(p^2_i)\} {(2\pi)^4 \delta^4(\sum p_i) \delta^8(\sum_i\lambda^i \eta_i)\over \left\langle 12\right\rangle \left\langle 23\right\rangle \left\langle 34\right\rangle \left\langle 41\right\rangle }
\label{W4}\ee
This answer for the amplitude is already cyclic symmetric.  In \cite{Kallosh:2009db} the answer analogous to eq. (\ref{W4}) was predicted for the light-cone computations (see eq. (2.19) in \N=8 supergravity case). Here we have demonstrated  the mechanism (in the  computation in \N=4 SYM case) which actually produces the  answer expected from equivalence theorem, as suggested  in Sec. 3 of \cite{Kallosh:2009db}. It is not surprising that in the tree approximation the equivalence theorem is confirmed, no anomalies would violate it. Still  it is satisfying to have a mechanism which converts the light-cone supergraph path integral into the one which produces the  covariant answers.

{\it Generating functional versus amplitudes}

To find all tree amplitudes, as shown in (\ref{total}),   from the generating functional (\ref{W4}) we have to look at the matrix element of $W^4(\phi_{in})$ between the vacuum and 4 outgoing states, with $ q_1, \eta_1, t^{a_1}$ etc. We have to perform a contraction between 4 fields $\phi_{in}(z_i)$ and 4 external states. If we agree to always contract $\phi_{in}(z_1)$ with $ q_1, \eta_1, t^{a_1}$ the remaining 3 superfields $\phi_{in}(z_2)$, $\phi_{in}(z_3)$, $\phi_{in}(z_4)$ may be contracted with the remaining 3 external states in a way which will result in permutation of $2,3,4$ as explained in (\ref{total}). Only one of these terms will give us a color ordered amplitude we are looking for, namely the first term in (\ref{6}). Therefore we conclude that 
\be
A_4^{\rm tree}(1,2,3,4)= g^2 {(2\pi)^4 \delta^4(\sum q_i) \delta^8(\sum_i\lambda^i (q)\eta_i)\over \left\langle 12\right\rangle \left\langle 23\right\rangle \left\langle 34\right\rangle \left\langle 41\right\rangle }
\label{A4}\ee


\subsubsection {Adding $ {\cal S}^3_\Delta $, $ {\cal S}^3_{\bar \Delta} $ and  $ {\cal S}^4 $ vertices, which cancel}

Now we have to look at the terms which were neglected so far in our computation of the 4-point on shell amplitude. A priory, one can expect two possibilities. The first one is that all supersymmetry/Lorentz symmetry breaking terms from $ {\cal S}_\Delta $, $ {\cal S}_{\bar \Delta} $ and  $ {\cal S}^4 $ vertices add together to something which is supersymmetry/Lorentz symmetry preserving. The second possibility is that they cancel. In such case we should see the mechanism of cancelation.

There is an extra $\bar \xi \xi $-dependent term in the fermionic, $\eta$-dependent part of the amplitude, and in the bosonic $\eta$-independent part of the amplitude. In all cases, there is at least one factor of $P^2$ which cancels the covariant propagator ${1/P^2}$. The corresponding ``shrinking tree''  graph is actually a contact term, since $\Box _x T(\phi_(x, \eta) \phi (x', \eta') \sim \delta^4(x-x')$. We present a corresponding graph symbolically in Fig. \ref{fig4}.
\begin{figure}[!h]
     \centering
     \subfigure{
\includegraphics[height=3.8cm]{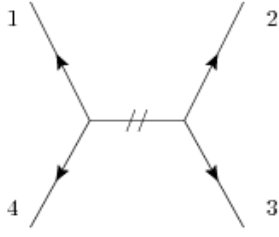}
 }
  \caption{A ``shrinking tree'' supergraph, which has a propagator $1/P^2$ canceled by $P^2$ term in the nominator.}
  \label{fig4}
  \end{figure}
The first place to look at is {\it the fermionic term} in eq. (\ref{ferm}) where now we have to keep the terms with  $(\xi \bar \xi)^P$
\be
 \delta^4 \left (\sum _{i=1}^4 \lambda^i_2 \eta_i  \right) \delta^4\left (\sum _{i=1}^4 \lambda^i_1 \eta_i -{\eta_2 ( \xi^P \bar \xi^P \bar \lambda^3)_1  - \eta_3 ( \xi^P \bar \xi^P \bar \lambda^2)_1 \over [23]} \right)
\label{ferm}\ee
 For on shell case with $p_2^2=p_3^2=0$ the new term proportional to $P^2$ is
\be
{\eta_2 \bar \lambda^3_{\dot 2}- \eta_3 \bar \lambda^2_{\dot 2} \over [23]}  (\bar \mu \mu)_{P 1 \dot 1} = {\eta_2 \bar \lambda_3- \eta_3 \bar \lambda_2\over [23]} \left ( {P^2\over P_+}\right)= {\psi_{32}\over [32]} \left ( {P^2\over P_+}\right)
\label{P2}\ee
Thus for the on shell case  the ``shrinking tree'' contribution to the 4-point amplitude is proportional to
\be
A^4_{contact} \sim  \delta^4 (\sum_{i=1}^4 \lambda_{2i} \eta_i)  \delta^4 (\psi_{32})   
\ee
Here we have recovered the same $\eta$-dependence as coming from the 4-vertex in $S^4$, see eqs.
(\ref{S4}), (\ref{S41}), (\ref{S42}).
Note that $P^2\neq 0$ in (\ref{P2}) since after $P$ integration  $P^2=(p_2+p_3)^2\neq 0$. The procedure of computations used here was to treat (\ref{BHT1}) as follows. The total expression is ${X(P)\over P^2}$. We present it as ${X(P^2=0)\over P^2} + {X(P)- X(P^2=0)\over P^2}$. The first term corresponds to ignoring terms proportional to $P^2$ in $X(P)$. Therefore the first terms has a pole in $P^2= (p_2+p_3)^2$, the second one does not have such a pole and corresponds to a contact term.

Thus we have found by computation that all dependence on Grassmann variables $\eta$ both in the quartic term in the action corresponding to the contact term in Fig. \ref{fig2} and the one in the 4-point shrinking tree supergraph in Fig. \ref{fig4} are exactly the same, $ F_{ij}(\lambda_2, \bar \lambda_{\dot 2 }, \eta) \equiv \delta^4 (\sum_{i=1}^4 \lambda_{2i} \eta_i)  \delta^4 (\eta_i\bar \lambda_{\dot 2 j} - \eta_j \bar \lambda_{\dot 2 i} )$. In both cases this $\eta$-dependent function is multiplied by a function of momenta: for the contact term let us call it  $B^{ij}_{\rm cont} (p)$ and for the the  shrinking tree let us call it  $B^{ij}_{\rm tree}(p)$.  For the quartic terms in the action the complete answer is given in eqs. (2.20)-(2.21). For the shrinking tree supergraph in Fig. \ref{fig4} we have  established the complete $\eta$-dependence, which we found the same as in the contact term in Fig. \ref{fig4}. As the result, the sum of these two supergraphs in Figs. \ref{fig2} and \ref{fig4} will be given by a function of $\eta$, $F_{ij}(\lambda_2, \bar \lambda_{\dot 2 }, \eta)$,  which is a common factor for both supergraphs,  times the sum of the  functions of momenta 
\be
A_4 ({\cal S}^3_\Delta ,  {\cal S}^3_{\bar \Delta},   {\cal S}^4 ) = F_{ij} (\lambda_2, \bar \lambda_{\dot 2 }, \eta) \left[ B^{ij}_{\rm cont} (p) + B^{ij}_{\rm tree}(p)\right]= \delta^4(Q_2) \delta^4 (\psi_{ij}) \left[ B^{ij}_{\rm cont} (p) + B^{ij}_{\rm tree}(p)\right]
\ee
The  supersymmetry generators act only on $\eta$-dependent factor, namely, the action of  8 kinematical supersymmetries is given by 
$
Q_{A 2}\equiv \sum_{i=1}^{4} \lambda_{2i}  \eta_{i A}$ and $ \bar Q_{\dot 2}^A \equiv \sum_{i=1}^4 \bar \lambda_{ \dot 2 i }{\partial \over \partial \eta_{Ai}} 
$.
Both $Q_{A 2}$ and $Q_{\dot 2}^A$ annihilate $ F_{ij}$, namely
$
Q_{A 2} F_{ij}= \bar Q_{\dot 2}^A F_{ij}=0
$.
However, both dynamical supersymmetries,  $Q_{A 1}$ and $Q_{\dot 1}^A$, do not annihilate $ F_{ij}$ as explained in Sec. 3.4. Moreover, it is known that the unique 4-point amplitude has the Lorentz covariant  dependence 
on $\eta$, it is given by 
\be
A_4 (\tilde {\cal S}^3  ) =  \delta^4 \sum_{i=1}^{4}( \lambda_{2i} \eta_i) \delta^4 \sum_{j=1}^{4} (\lambda_{1j}
\eta_j) B(p)
=
 \delta^4(Q_2) \delta^4 (Q_1) B(p)= \delta^8 (Q_\alpha) B(p)
\ee
This means that $A_4 ({\cal S}^3_\Delta ,  {\cal S}^3_{\bar \Delta},   {\cal S}^4 )$ must vanish.
Thus the role of ${\cal S}^4$ is to cancel the supersymmetry  breaking ``shrinking tree'' contribution to the 4-point amplitude, shown in Fig. 6.

The second source of the {\it  bosonic terms}  with $ {\cal S}_\Delta $, $ {\cal S}_{\bar \Delta} $ vertices comes from the correction to the eq. (\ref{1}) which with the account of the omitted $\bar \xi \xi$-dependent terms is given by:
\be
\left\langle 1P\right\rangle [P3]= \left\langle 12\right\rangle [23] - \lambda^1 (\xi \bar \xi)_P \bar \lambda^3 \, , \qquad
\left\langle 4P\right\rangle [P2]= \left\langle 43\right\rangle [32] - \lambda^4 (\xi \bar \xi)_P \bar \lambda^2
\label{2'}\ee
In the 4-point amplitude
\be
 {[23]^4 \delta^4 \left (\sum_{i=1}^4 p_i\right) \delta^8 \left (\sum_{i=1}^4 \lambda_{\alpha i} \eta_i\right) \over  \left\langle 23\right\rangle [23] \left\langle 41\right\rangle [23]( \left\langle 12\right\rangle [23]-\Delta_{13})
( \left\langle 34\right\rangle  [23] -\Delta_{42})} = {\delta^4 \left (\sum_{i=1}^4 p_i\right) \delta^8 \left (\sum_{i=1}^4 \lambda_{\alpha i} \eta_i\right)\over \left\langle 12\right\rangle \left\langle 23\right\rangle \left\langle 34\right\rangle \left\langle 41\right\rangle -\Delta }
\label{BHTanswerii}\ee
the extra terms are Lorentz non-covariant terms $\Delta^{13}= \lambda^1 (\xi \bar \xi)_P \bar \lambda^3$ and $\Delta^{42}= \lambda^4 (\xi \bar \xi)_P \bar \lambda^2$.

The extra   bosonic terms come with $\eta$-dependence and momentum dependence of the form which is manifestly supersymmetric,  namely, with $\delta^4(P) \delta^8 (Q_\alpha)$.

One finds that the terms originating from $\Delta^{13}$ and $\Delta^{42}$ lead to the expression for $\Delta$ which is proportional to
\be
\Delta(p_{i\bot} , p_{i+}) \sim {\left\langle 12\right\rangle \over (\lambda^1 \bar \lambda^3)_{2\dot 2}} + {\left\langle 34\right\rangle \over (\lambda^4 \bar \lambda^2)_{2\dot 2}} - {\left\langle 23\right\rangle \over (\lambda^2 \bar \lambda^2+ \lambda^3 \bar \lambda^3 )_{2\dot 2}}
\ee
Here we are using the spinors $\lambda$ and $\mu$ as they come from the light-cone gauge, given in Appendix. $\Delta$  is linear in  $p_{i\bot}$, since the expression is linear in angular brackets and all $\lambda_2$ and $\bar \lambda_{\dot 2}$ depend only on $p_+$. A choice of the frame, $\Delta(p_{i\bot} , p_{i+})=0$ removes the Lorentz non-covariant terms from the amplitude.

In conclusion of this section: adding supergraphs with $ {\cal S}_\Delta $, $ {\cal S}_{\bar \Delta} $ and  $ {\cal S}^4 $ vertices did not affect the answer for the 4-point on shell amplitude since all additions cancel.

\subsection{2 MHV or two $\overline { \text{MHV}}  $  vertices}

Now we compute the  supergraph in Fig. 5 with 2 MHV vertices. 
\begin{figure}[!h]
     \centering
     \subfigure{
\includegraphics[height=3.5cm]{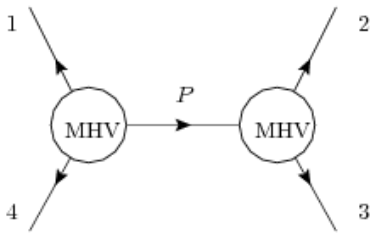}
 }
    \label{fig5}
  \caption{Two  MHV vertices}
  \end{figure}
  We find that for this computation it is simpler instead of the spinor product
expression to use the vertex in the form of eq.  (\ref{eq:s3step2})  for both vertices which are connected by a propagator 
$
\delta^{4}(\eta_{P}+\eta_{-P})/P^{2}$. After integrating out the Grassmann variables the answer is proportional to
\begin{equation}
\frac{\left\langle 12 \right\rangle   \left\langle 34 \right\rangle}{\left\langle 34\right\rangle \left[34\right]}\, .\label{eq:fact1}\end{equation}
This expression has to be made cyclic symmetric for the description of the partial color ordered amplitude, thus we add
 $\frac{\left\langle 23 \right\rangle   \left\langle 41 \right\rangle}{\left\langle 41\right\rangle \left[41\right]}$. The result vanishes  due to
 momentum conservation $\sum_{i=1}^4\lambda_{i}\bar{\lambda}_{i}=0$.
\begin{equation}
\left\langle 12\right\rangle \left[41\right]+\left\langle 23\right\rangle \left[34\right]=0\end{equation}
In the case of two $\overline { \text{MHV}}  $ vertices one finds the same situation, the contribution to the 
cyclicaly symmetric  partial color ordered amplitude vanishes.

{\it Grassmann degree argument}

Following \cite{Drummond:2008cr} we may employ the Grassmann degree argument to prove that the 4-point light-cone amplitude with 2 MHV or  two $\overline { \text{MHV}}  $  vertices must vanish. Indeed, the resulting 4-point amplitude MHV has  Grassmann degree 8. However, in the case of 2 MHV vertices one finds the Grassmann degree
\be
2\times 8-4= 12
\ee
Each MHV gave 8 and one propagator gave -4. 12 is not possible for the 4-point MHV amplitude, so the supergraphs on Fig. 7 have to cancel, which we also checked directly.
The 4-point amplitude in the case of 2 $\overline { \text{MHV}}  $ vertices has the Grassmann degree
\be
2\times 4-4= 4
\ee
and again, this is not equal to 8. The amplitude should vanish, as we have seen above. 

Only the case of one 3-MHV vertex and one 3-$\overline { \text{MHV}}  $ vertex has a correct Grassmann degree $8+4-4=8$ for the 4-point amplitude.


\section{Background field method for the tree level light-cone superfields}

It is convenient to use the background field method \cite{DeWitt:1967ub} to present the compact total answer for all light-cone tree amplitudes in terms of the background field $\varphi[\phi_{\rm in}]$ which solves the classical field equations in presence of the external source $J[\phi_{\rm in}] $.   To compute the path integral in eq. (\ref{pathintYM}) we expand  it around a stationary point. 

\subsection{$ \phi^3$ example}

We remind the procedure for the case of a simple action of a  scalar field with cubic interaction. The generating functional for the connected Green functions is
\be
e^{i Z[J]}  =  \int d\phi~  \exp {i \left (S[\phi]  +  J_i \phi^i \right) }\ .
\ee
where
$S[\phi]={1\over 2} \phi^i S_{,ij} \phi^j +{1\over 3} S_{,ijk} \phi^i \phi^j \phi^k $. In condensed DeWitt's notation \cite{DeWitt:1967ub} a summation over $i$ includes the integration over $d^4 x$.
The $S$-matrix is obtained via LSZ reduction which corresponds to replacing the external source term 
$J_i \phi^i $ by $ =\phi_{\rm in} ^i {\vec S}_{,ij} \phi^j$, which leads to 
\be
 \exp {i W[\phi_{\rm in} ]}=  \int d\phi~  \exp {i \left (S[\phi]  -  \phi_{\rm in} ^i {\vec S}_{,ij} \phi^j \right) }\ .
\label{pathintSC}\ee
Here $\phi_{\rm in}$ is a free field satisfying equation $S_{ij} \phi^j_{\rm in}=0$.
The stationary point  $\varphi^i$ is given by the equation
\be
 S_{,i} -  S_{,ij}\phi_{\rm in}^j= S_{,ij}(\varphi-\phi_{\rm in})^j + S_{,ijk}  \varphi^j \varphi^k=0
\ee
The Green function is defined as an inverse to the differential operator of the quadratic terms in the action $S_{,ij} G^{jk}= -\delta_i^k$.   Here $\delta_i^k$ includes also $\delta^4(x-x')$ since the Green function is non-local. The stationary point of the path integral  defines the background field $\varphi[\phi_{\rm in}]$:
\be
\varphi^i = \phi_{\rm in}^i + G^{ij}  S_{,jkl}  \varphi^j \varphi^l  
\label{classical}\ee
 This equation has  an iterative solution
\be
\varphi ^i[\phi_{\rm in}]= \phi_{in}^i +  G^{ij}   \sum_{n=2} ^\infty t_{j i_1...i^n} \phi_{\rm in}^{i_1}... \phi_{\rm in}^{i_n}
\label{iter}\ee
which shows the decomposition of the background field into a tree-graph structure with any number of legs. The value of the exponent of the integral (\ref{pathintSC}) at the stationary point is
\be
 W [\varphi, \phi_{ \rm in} ]_{\rm tree} =   {1\over 2} (\varphi -\phi_{\rm in})^i S_{,ij} (\varphi -\phi_{\rm in})^j +{1\over 3} S_{,ijk} \varphi^i \varphi^j \varphi^k
\ee
where we used the fact that $ \, \phi_{\rm in}^i S_{,ij} \phi_{\rm in}^j=0$. 

We may  rewrite it  in the form where it depends only on the background field $\varphi[\phi_{in}]$, using (\ref{classical}) 
\be\boxed{
 W [\varphi[\phi_{\rm in}] ]_{\rm tree} =  - {1\over 2}   S_{,ikl}  \phi^k \phi^l \,  G^{ij} \, S_{,jnp}  \phi^n \phi^p   +{1\over 3} S_{,ijk} \phi^i \phi^j \phi^k}
\ee
When one inserts the iterative solution of  (\ref{classical})  for $\phi$ in terms of $\phi_{in}$ as shown in (\ref{iter}),  one finds all tree diagrams of the theory. 

For example the 4-point amplitude comes from two sources: from the first term we get
\be
 - {1\over 2}   S_{,ikl}  \phi^k_{\rm in} \phi^l_{\rm in} \,  G^{ij} \, S_{,jnp}  \phi^n_{\rm in} \phi^p_{\rm in}\ee
From the second one one finds the same terms with the coefficient $+1$.
The total contribution to 4-point generating function is
\be
 W^4_{\rm tree} =   {1\over 2}   S_{,ikl}  \phi^k_{\rm in} \phi^l_{\rm in} \,  G^{ij} \, S_{,jnp}  \phi^n_{\rm in} \phi^p_{\rm in}
\label{W4}\ee
We may present it in the form
\be
 W^4_{\rm tree} =  W_{klnp}   \phi^k_{\rm in} \phi^l_{\rm in} \phi^n_{\rm in} \phi^p_{\rm in}
\label{W4'} \ee
It connects two 3-point vertices by a propagator. The $S$-matrix element can be computed when this expression is inserted between physical states with particular momenta etc.

For the 5-point generating function one finds, by keeping one more power of $\phi_{\rm in}$ in the expansion,
\be
W^5_{\rm tree} =   - S_{,ikl}  \phi^k_{\rm in} \phi^l_{\rm in} \,  G^{ij} \, S_{,jnp}  \phi^n_{\rm in} \, G^{pm} \, S_{,mqr}  \phi^q_{\rm in} \phi^r_{\rm in}
\label{W5tree}\ee
This is a graph which connects three 3-vertices by two propagators. Or, equivalently, it may be understood as a 4-point amplitude in which one off-shell leg was replaced by the second term in (\ref{classical}).
The corresponding recursion relations remind the ones, derived for the tree-level gluons in \cite{Berends:1987me}.

\subsection{Application to \N=4  light-cone supergraphs}
In the application to the \N=4  light-cone supergraphs we propose to use the Feynman rules as explained above, in the computation of the 4-point amplitude (\ref{A4}). In the background field method the quadratic part of the action will define $S_{,ij}$ and its inverse $G^{ij}$, whereas the cubic part will define $S_{,ijk}$. We will not involve the quartic vertex, however, we will have to perform the computation as explained in Sec. 5 where the shrinking tree supergraphs are also neglected. 
One can now check that the 4-point generating function, proposed in (\ref{W4}), (\ref{W4'}) will correspond to the computation leading to 4-point amplitude (\ref{A4}). 

For the computation of the 5-point generating function it is helpful to use the expression given in the background field method in (\ref{W5tree}). It means that we have to take a 4-point amplitude and contract  one of its legs to the 3-point amplitude.  The 4-point amplitude is MHV, however, the 3-point amplitude may be either MHV or $\overline { \text{MHV}}  $. In case, it is $\overline { \text{MHV}}  $, we get  a 5-point amplitude with the Grassmann degree 8+4-4=8 and we get the MHV 5-point amplitude. This is the case closely related to the computation we did for the 4-point amplitude in Sec. 5 where we contracted a 3-point MHV vertex with the 3-point $\overline { \text{MHV}}  $. For the 5-point case the computation is almost the same as for the 4-point case, we will present it below. For the NMHV 5-point amplitude one has to contract, according to (\ref{W5tree}), the 4-point MHV amplitude with the 3-point MHV, the Grassmann degree will be 8+8-4=12, which is required for the NMHV amplitude. The details of the computation are below.

\section{ Computation of the 5-point tree supergraphs}

\subsection{MHV case}

For the supergraph with an MHV 4-vertex with 1, 5, 4 , P and the  ${\overline{\text{MHV}}}$ vertex with 2, 3, P  in Fig. 1 we have to integrate over $P$ and $\eta_P$  the following expression
\be
{\delta^4(p_1+p_5 + p_4+P) \delta^8 (\lambda^1\eta_1+ \lambda^5\eta_5+ \lambda^4 \eta_4 + \lambda^P \eta_P ) \delta^4(-P+p_2+p_3)  \delta^4( \eta_P [23] + \eta_2[3P] +\eta_3 [P2])\over P^2 \left\langle 51\right\rangle  \left\langle 45\right\rangle [23] \left\langle 1P\right\rangle \left\langle P4\right\rangle [P2] [3P] }
\label{BHT5}\ee
We now use the fact that $\eta_P=  {\eta_2[P3] -\eta_3 [P2]\over [23]}
$.
We insert this expression into $ \delta^8 (\lambda^1\eta_1+  \lambda^5\eta_5 + \lambda^4 \eta_4 + \lambda^P \eta_P)$
and find
\be
 \delta^8 \left (\sum _{i=1}^5 \lambda^i \eta_i + {\eta_2 ( \lambda^P [P3] -\lambda^2 [23]) - \eta_3 ( \lambda^P  [P2] - \lambda^3 [32])  \over [23]} \right)
\ee
 We rearrange the argument of the fermionic $\delta$-function as follows
\be
 \delta^8 \left (\sum _{i=1}^5 \lambda^i \eta_i + {\eta_2 ( \lambda^P \bar \lambda^P -\lambda^2 \bar \lambda^2 -  \lambda^3 \bar \lambda^3) \bar \lambda^3 - \eta_3 ( \lambda^P \bar \lambda^P -\lambda^2 \bar \lambda^2 -  \lambda^3 \bar \lambda^3) \bar \lambda^2 \over [23]} \right)
\label{D}\ee
According to the prescription (i)  in sec. 5.1 we use 
$( \lambda \bar\lambda)^2+( \lambda \bar\lambda)^3-( \lambda \bar\lambda)^P=0$, the fermionic $\delta$-function (\ref{D}) becomes
\be
 \delta^8 \left (\sum _{i=1}^5 \lambda^i \eta_i  \right)
\ee
Using the same steps as before we find
\be
 {[23]^4\over  \left\langle 23\right\rangle [23] \left\langle 51\right\rangle  \left\langle 45\right\rangle  [23] \left\langle 12\right\rangle \left\langle 34\right\rangle [23] [23] } = {1\over \left\langle 12\right\rangle \left\langle 23\right\rangle \left\langle 34\right\rangle \left\langle 45\right\rangle  \left\langle 51\right\rangle  }
\label{BHTanswer}\ee
This reproduces the correct generating function for the 5-point on shell amplitude
\be
W^5(\phi_{\rm in})=  g^3 tr \int \prod_{i=1}^5 \{d^8 z_i \phi_{\rm in}(z_i) \delta(p^2_i)\} {(2\pi)^4 \delta^4(\sum p_i) \delta^8(\sum_i\lambda^i \eta_i)\over \left\langle 12\right\rangle \left\langle 23\right\rangle \left\langle 34\right\rangle \left\langle 45\right\rangle  \left\langle 51\right\rangle }
\label{W5'}\ee

\subsection{NMHV case}

Here we have to contract a covariant part of the 3-vertex (1P5) with the 4-point amplitude (234P) via a propagator $1/P^2$. The relevant expression was also studied in \cite{Brandhuber:2008pf} in the context of recursion relations. It has  a denominator and a nominator. We start with the denominator
\be
{1\over P_{15}^2}{1\over \left\langle 1P\right\rangle \left\langle P5\right\rangle \left\langle 51\right\rangle
\left\langle P2\right\rangle \left\langle 23\right\rangle \left\langle 34\right\rangle \left\langle 4P\right\rangle} =
{1\over \prod_{i=1}^5 \left [ii+1\right ]} {\left[34\right]^4\over \left\langle 15\right\rangle^4 \left\langle 2P\right\rangle^4}
\ee
The Grassmann part of the nominator is
\be
\int d^4 \eta_P  \, \delta^8( \lambda^1 \eta_1 +\lambda^5 \eta_5 +\lambda^P \eta_P) \, \delta^8( \lambda^2 \eta_2 +\lambda^3 \eta_3 + \lambda^4 \eta_4 -\lambda^P \eta_P)
\ee
Here we can first rewrite it as 
\be
\delta^8( \lambda^1 \eta_1 + \lambda^2 \eta_2 + \lambda^3 \eta_3 +  \lambda^4 \eta_4+ \lambda^5 \eta_5 ) \,  \int d^4 \eta_P  \, \delta^8( \lambda^2 \eta_2 +\lambda^3 \eta_3 + \lambda^4 \eta_4 -\lambda^P \eta_P)
\label{int}\ee
Now we observe that
\be
\delta^8( \lambda^2 \eta_2 +\lambda^3 \eta_3 + \lambda^4 \eta_4 -\lambda^P \eta_P)
= \left\langle 2P\right\rangle^4 \delta^4 \left ( \eta_P- {1\over \left\langle 2P\right\rangle^4} \sum_{k=3,4} \left\langle 2k\right\rangle \eta_k\right ) \delta^4 \left (\eta_2 + {1\over \left\langle P2\right\rangle} \sum_{k=3,4} \left\langle Pk\right\rangle \eta_k\right )
\ee
We now perform the $\eta_P$ integration
and find that the nominator becomes
\be
\left\langle 2P\right\rangle^4 \delta^8 \left (\sum_{i=1}^5{ \lambda^i \eta_i}\right)  \delta^4 \left (\eta_2 + {1\over \left\langle P2\right\rangle}  (\left\langle P3\right\rangle \eta_3 +  \left\langle P4\right\rangle \eta_4 )\right )
\ee
To combine the nominator with the denominator we have also to take into account that the properties of the 4-point amplitude like $ \left\langle 3P\right\rangle^4/ \left\langle 2P\right\rangle^4 = \left[24\right]^4/\left[34\right]^4$. The result for the 5-point tree supergraph NMHV amplitude is
\be
W^5(\phi_{in})=  tr \int \prod_{i=1}^5 \{d^8 z_i \phi(z_i) \delta(p^2_i)\} {(2\pi)^4 \delta^4(\sum p_i) \delta^8(\sum_{i=1}^5 \lambda^i \eta_i)\delta^4 (\eta_2 [34] +\eta_3[42] + \eta_4[ 23])   \over \left\langle 15\right\rangle^4  \prod_{i=1}^5 \left [ii+1\right ]}
\label{W5}\ee
This agrees with the 5-point NMHV amplitude given in eq. (2.25) of \cite{Drummond:2008bq}. However, it was derived there from the known expression for the MHV amplitude via the transition to the anti-chiral basis $\eta \rightarrow \bar \eta$ variables and back. Here we have computed the 5-point NMHV amplitude via the rules of the background field method version of new path integral.

\section{Computation of the MHV n-point  color ordered amplitude}
This case is a simple generalization of the 4-point and 5-point MHV cases above: the $(n-1)$-point MHV amplitude is contracted with the 3-point ${\overline{\text{MHV}}}$ vertex. The $(n-1)$-point MHV amplitude comes with the $\delta$-function of the form
$\delta^4(p_1+p_n+... +P) \delta^8 (\lambda^1\eta_1+ \lambda^n\eta_n+ ... + \lambda^P \eta_P )$. According to rules above this $\delta$-function, when multiplied on the 3-point ${\overline{\text{MHV}}}$ vertex $\delta$-function, is resolved to produce the required $\delta^4( \sum _{i=1}^5 p_i) 
 \delta^8 \left (\sum _{i=1}^5 \lambda^i \eta_i  \right). $ The extra angular brackets defining the $(n-1)$-point MHV amplitude are exactly the ones which produce the complete set of angular brackets for the  $n$-point MHV amplitude, as it was shown in detail for the 4- and 5-point cases.
 
 The corresponding recursion relations are very much in spirit of the ones, derived for the tree-level gluons in \cite{Berends:1987me} where, in particular, the MHV amplitudes were shown to solve the recursion relation. The same for the light-cone superfield amplitudes, in MHV case they solve the recursion relation associated with (\ref{iter}), (\ref{classical}). Moreover, since we deal with the light-cone scalar superfields, we do not have to deal with the complicated kinematics, we effectively take an  $(n-1)$-point MHV  amplitude with one leg off shell and it is replaced by a second term in (\ref{iter}), which upon integration over $P, \eta_P$ produces the $n$-point MHV  amplitude. The result is

 \be
W^n_{\rm MHV} (\phi_{\rm in})=  g^{n-2} tr \int \prod_{i=1}^n \{d^8 z_i \phi_{\rm in}(z_i) \delta(p^2_i)\} {(2\pi)^4 \delta^4(\sum p_i) \delta^8(\sum_i\lambda^i \eta_i)\over \left\langle 12\right\rangle \left\langle 23\right\rangle \left\langle 34\right\rangle \left\langle 45\right\rangle ...  \left\langle n1\right\rangle }
\label{W5'}\ee
 
 \section{ Computation of the 6-point  NMHV tree supergraphs}

Here we will not go into details of the complete 6-point amplitude as given by the background functional method, this will require more studies in the future. However, we will show here that a contraction of the two 4-point MHV amplitudes, using the rules of the new path integral,  does  produce a correct 6-point NMHV amplitude, which has a correct Grassmann degree 8+8-4=12.  It is interesting here that we are not using a the complexification of momenta and shifts which are usually used for the recursion relations, as for example it was done in the computations of the all tree-level amplitudes in \cite{Drummond:2008cr}. Reading  \cite{Drummond:2008cr} one gets an impression that  the factors of the type ${1\over \left\langle 1|  p_2+p_3 | 4 \right ]} $ come from the momentum shifts, and therefore  it is difficult to see how they could arise from Feynman rules without shifts\footnote{We are grateful to J. Kaplan for a discussion of this issue}. In our computation one can see that such terms originate from the the Grassmann integration. 

Thus we  contract a  4-point amplitude (123P) with the 4-point amplitude (456P) via a propagator $1/P^2$,
\be
\int {d^4 P d^4\eta_P\over P^2}  {\delta^4(p_1+p_2+p_3+P)  \delta^8(q_1+q_2+q_3+q_P)\over \left\langle 12\right\rangle \left\langle 23\right\rangle \left\langle 3P\right\rangle
\left\langle P1\right\rangle}{\delta^4(p_4+p_5+p_6-P)  \delta^8(q_4+q_5+q_6-q_P)\over  \left\langle P4\right\rangle \left\langle 45\right\rangle \left\langle 56\right\rangle
 \left\langle 6P\right\rangle} \ee
where $q_i=\lambda^i \eta_i$.
The integration over $P$ leads to  $P=-(p_1+p_2+p_3)=p_4+p_5+p_6$. Meanwhile, for the 4-vertices the rule is to use the expressions where $\delta^4(p_1+p_2+p_3+P) $ is replaced by $\delta^4((\lambda \bar \lambda)_1+(\lambda \bar \lambda)_2+(\lambda \bar \lambda)_3+(\lambda \bar \lambda)_P) $  This means that the 4-vertex is taken at all momenta on shell, namely $p_1^2=p_2^2=p_3^2=P^2=0$. And  the same for the other 4-vertex.

First we use the momentum conservation for the second 4-vertex $\delta^4((\lambda \bar \lambda)_4+(\lambda \bar \lambda)_5+(\lambda \bar \lambda)_6-(\lambda \bar \lambda)_P) $ and find that
\be
{1\over 
 \left\langle P4\right\rangle \left\langle 45\right\rangle \left\langle 56\right\rangle
 \left\langle 6P\right\rangle} =
{[56]^4\over [P4][45][56][6P] \left\langle 4P\right\rangle^4}
\ee
Now we reorganize the second Grassmann part
\be
\delta^8( \lambda^4 \eta_4 +\lambda^5 \eta_5 + \lambda^6 \eta_6-\lambda^P \eta_P)
= \left\langle 4P\right\rangle^4 \delta^4 \left (   \sum_{k=5,6} {
\left\langle 4k\right\rangle \over \left\langle 4P\right\rangle}\eta_k
 -\eta_P\right ) \delta^4 \left (\eta_4 +   \sum_{k=5,6} {
\left\langle k P\right\rangle \over \left\langle 4P\right\rangle}\eta_k \right)
\ee
We now perform the $P$ and $\eta_P$ integration
and find 
\be
 \delta^4\left (\sum_{i=1}^6 p_i\right ) \delta^8 \left (\sum_{i=1}^6{ \lambda^i \eta_i}\right)  \delta^4 \left (\eta_4 + {\left\langle P5\right\rangle\over \left\langle P4\right\rangle}   \eta_5 +  {\left\langle P6\right\rangle\over \left\langle P4\right\rangle} \eta_6 )\right ){[56]^4\over P^2 [P4][45][56][6P]  \left\langle 12\right\rangle \left\langle 23\right\rangle \left\langle 3P\right\rangle
\left\langle P1\right\rangle}
\ee
We have  to take into account that the properties of the on shell 4-point amplitude like ${ \left\langle P4\right\rangle \over [56]} = {\left\langle 5P\right\rangle\over [46]}$. This leads to the following 
\be
 \delta^4\left (\sum_{i=1}^6 p_i\right ) \delta^8 \left (\sum_{i=1}^6{ \lambda^i \eta_i}\right)  \delta^4 \left (\eta_4 [56]+    \eta_5 [64]+  \eta_6[45] )\right ){1\over P^2  [45][56]  \left\langle 1P\right\rangle [P4]  \left\langle 3P\right\rangle [P6]  \left\langle 12\right\rangle \left\langle 23\right\rangle 
}
\ee
Note that 
\be
 \left\langle 1P\right\rangle [P4]=  \left\langle 1| P | 4 \right ]= \left\langle 1| p_2+p_3 | 4 \right ]
\label{1}\ee
and
\be
 \left\langle 3P\right\rangle [P6]=  \left\langle 3| P | 6 \right ]= \left\langle 3| p_4+p_5 | 6 \right ]= \left\langle 3| p_1+p_2 | 6 \right ]
\label{2}\ee
Introducing the standard notation we can represent the NMHV 6-point amplitude as
\be
W^6(\phi_{in})\sim  tr \int  \prod_{i=1}^6  d^8 z_i \phi(z_i) \delta(p^2_i)  \, { (2\pi)^4 \delta^4(\sum p_i) \delta^8(\sum \lambda^i \eta_i)  \over \left\langle 12 \right\rangle   \left\langle 23 \right\rangle  \left\langle 34 \right\rangle  \left\langle 45 \right\rangle  \left\langle 56 \right\rangle  \left\langle 61 \right\rangle}  ( R_{146} + \rm cyclic )
\label{W6}\ee
 Here
\be
R_{146} = { \left\langle 34 \right\rangle  \left\langle 45 \right\rangle  \left\langle 56 \right\rangle  \left\langle 61 \right\rangle\over P^2 [45][56] \left\langle 1| p_2+p_3 | 4 \right ] \left\langle 3| p_1+p_2 | 6 \right ]} \, 
 \delta^4 (\eta_4 [56] +\eta_5[64] + \eta_6[ 45]) 
\label{146}\ee
and ($R_{146}+ $ cyclic) means $R_{146}+ R_{251} +R_{362} + R_{413}+ R_{524} + R_{635}$.
This agrees with the 6-point NMHV amplitude given in  \cite{Korchemsky:2009hm}.

Note that various contact terms proportional to $P^2$ have beed neglected according to ``rules'' established for 4-point amplitude where such terms are cancelled by the contribution from the original 4-point vertex, which was also neglected. This computation is rather interesting since it gives an evidence that starting from the unitary light-cone superfield path integral for \N=4 super-Yang-Mills theory, one can develop a  manifestly supersymmetric unitarity cut method.

\section{Conclusion and Discussion}

We have reorganized the light-cone supergraph path integral for \N=4 Yang-Mills theory using the Lorentz covariant spinor helicity formalism. The Feynman rules in a Fourier superspace  produce the on shell amplitudes which are split into a part (i) which  at  every step of computation preserves the unbroken 16 supersymmetries, 8 kinematical and 8 dynamical ones, as well as a Lorentz symmetry. The part (ii) has a contribution from vertices which individually  break dynamical supersymmetry and Lorentz symmetry, they preserve only a kinematical supersymmetry.

Given this split of the answer into a covariant part (i) and the non-covariant part (ii) one would like to find out  if the sum of the supergraphs in part (ii) cancels or combines into an additional covariant part of the answer.

\begin{figure}[!h]
     \centering
     \includegraphics[height=3cm]{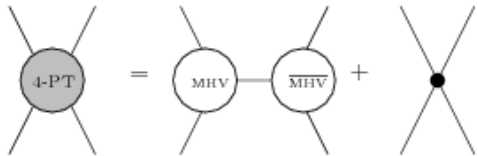}
   \caption{The total 4-point tree amplitude is given by the sum of two  supergraphs:
a tree with the 3-vertices  MHV   and ${\overline{\text{MHV}}}$, and a contact term. Each of these two supergraphs breaks dynamical supersymmetry and Lorentz  symmetry.}
  \label{fig6}
  \end{figure}

According to the path integral the total 4-point tree amplitude is given by the sum of the  supergraphs presented in Fig. 9.
It consists of the tree supergraphs with the 3-vertices  (\ref{V}) and (\ref{barV}) and contact terms with
4-point vertices  (\ref{S41}) and (\ref{S42}).

For the tree level 4-point amplitude we have computed the (i) part of the supergraphs and analyzed the (ii) part.   The MHV - ${\overline{\text{MHV}}}$ supergraph on the rhs of the Fig. \ref{fig6}  consists of two contributions described in details in Sec. 5. One part is supersymmetric and Lorentz covariant and employs only $\tilde {\cal S}_{MHV}^3$ and $\tilde {\cal S}^3_{\overline{\text{MHV}}}$ vertices, see Sec.  5.1.1. This is the (i) part of the computations. The other contribution is what we call a ``shrinking tree'' contribution, see Sec. 5.1.2.  Thus, together the 4-point amplitude has 3 contributions, see  Fig 10: a manifestly supersymmetric one, (i) part of the amplitude, the first term on the rhs of Fig. 10, and part (ii):  two terms breaking supersymmetry: a  ``shrinking tree'' contribution, the second  term in Fig. 10, which is a leftover from the MHV - ${\overline{\text{MHV}}}$ supergraph after the covariant part (i) was taken out of it,  and a contact term,  the third term in Fig. 10.

\begin{figure}[!h]
     \centering
\includegraphics[height=2.5cm]{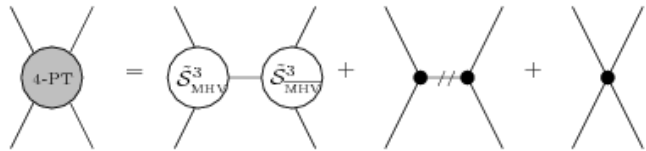}
   \caption{The MHV - ${\overline{\text{MHV}}}$ supergraph in the Fig. \ref{fig6} is split into a supersymmetric part,
   with $\tilde {\cal S}_{MHV}^3$ -  $\tilde {\cal S}^3_{\overline{\text{MHV}}}$ vertices,
   the first term on the rhs  and a ``shrinking tree'' supergrapgh, the second term on the rhs. Together with the third contact term on the rhs, the second term forms an (ii) part of the answer, which vanishes for the on shell 4-point amplitude. Only the first term in the rhs survives and gives the correct answer.}
  \label{fig7}
  \end{figure}

One of the  interesting and unexpected results of the reformulation of the path integral proposed in this paper is the
cancellation of contact terms in the 4-point tree amplitude.
There are two sources of the contact terms in the 4-point amplitude. There is a contribution from   the  ${\cal S}^3_\Delta , {\cal S}^3 _{\bar \Delta}$ vertices in a ``shrinking tree'' supergraph, and  from the ${\cal S}^4$ vertex.

In the case of the tree 4-point amplitude we can clearly see that these two sources of the contact terms must cancel. The answer in eq. (\ref{W4}) coming from the simple and manifestly supersymmetric vertices $\tilde {\cal S}_{MHV}^3$ and $\tilde {\cal S}^3_{\overline{\text{MHV}}}$ is already a correct one. It has 16 supersymmetries and is Lorentz covariant. It is unique, up to an overall factor.

Both contact terms, the second and the third one in Fig. \ref{fig7},  break dynamical supersymmetry and Lorentz symmetry. They both depend on the Grassmann variable $\eta$'s via some combination of the functions $\delta^4(\psi_{ij})$, breaking dynamical supersymmetry. Therefore adding two contributions with the same dependence on $\eta$'s will not convert this dependence into the one we need for unbroken dynamical supersymmetry, namely $\delta^4(Q_1)$. Thus, the contributions from the ``shrinking trees'' and from the 4-point vertex can't combine into an expression with dynamical supersymmetry unbroken, they can only cancel. So, at least at the level of the 4-point tree amplitude, which we studied so far in the framework of the new path integral, we do not need to know the detailed form of all complicated terms which break the dynamical supersymmetry, they cancel.

In this respect it it interesting to ask: why the path integral provides the last two terms in the rhs of Fig. \ref{fig7},   which cancel anyway. The surviving contribution, the first term in Fig. \ref{fig7}, has an interesting feature associated with the
recursion relations \cite{Britto:2004ap}, \cite{Brandhuber:2008pf}, \cite{ArkaniHamed:2008gz}   for the  on shell  amplitudes. In our path integral the surviving first term on the rhs of Fig. \ref{fig7} has vertices $\tilde {\cal S}_{MHV}^3$ and $\tilde {\cal S}^3_{\overline{\text{MHV}}}$ which are taken at the residue of the pole, at $P^2=0$. In the situation with the recursion relations this means that the vertex vanishes, unless  some of the outgoing momenta in the vertex are complexified. Meanwhile, in the path integral the tree supergrapgh with MHV - ${\overline{\text{MHV}}}$ vertices, the first term in the rhs of Fig. \ref{fig6},  is given by the expression where the vertices are not taken at the residue of the pole. So the total MHV - ${\overline{\text{MHV}}}$ tree graph puts no restriction on the external momenta,  therefore it is not necessary to complexify the momenta.
However, it turns out that the total answer for the tree MHV - ${\overline{\text{MHV}}}$ supergraph can be split into two terms, one which correspond to the vertices at the residue at the pole, $\tilde {\cal S}_{MHV}^3$ - $\tilde {\cal S}^3_{\overline{\text{MHV}}}$ and the other one, the ``shrinking tree'' graph. Besides, there is also a contact term, the last in the rhs of Fig. \ref{fig7}. It cancels the  ``shrinking tree'' graph and the answer is only the first $\tilde {\cal S}_{MHV}^3$ - $\tilde {\cal S}^3_{\overline{\text{MHV}}}$ term in Fig. \ref{fig7}. 

If the mechanism of cancellation of the complicated non-supersymmetric contributions to the on shell amplitudes would work also for more external legs and more loops, it would simplify the computations significantly. 

For \N=8 supergravity the analogous reorganization of the light-cone path integral would be extremely desirable. It is likely that at the tree level all steps which were performed for \N=4 supersymmetric Yang-Mills theory will also work for \N=8 supergravity. The action is known only up to a cubic order  in superfields.  If the mechanism of compensation of contact terms  with broken dynamical supersymmetry, which we have found in \N=4 SYM,  would work in 
\N=8 supergravity, it would mean that all  complicated contact terms in the action are simply designed to remove the supersymmetry breaking terms from the graphs with cubic vertices only. In such case everything may be simplified, which would  make this approach to general analysis and actual computations quite efficient.

\section*{Acknowledgments}

We are grateful to Zvi Bern, Lars Brink, John Josef Carrasco, Jared Kaplan, Paul Mansfield, Bengt  Nilsson, Tomas Rube and Ed Witten for the stimulating discussions.   This work is supported by the NSF grant 0756174 and by Stanford Institute of Theoretical Physics.

\appendix

\section{ Notation}

In this paper we adopt a shorthand notation for light-cone coordinates
\begin{equation}
x_{\pm}=(x_{0}\pm x_{3})/\sqrt{2},\end{equation}
where the transverse components are given by
\begin{equation}
x_{\bot}=(x_{1}+ix_{2})/\sqrt{2},\,\bar{x}_{\bot}=(x_{1}-ix_{2})/\sqrt{2}.\end{equation}
In these coordinates the flat metric is off-diagonal. The scalar product
of two 4-vectors $x$ and $y$ reads
\begin{equation}
x\cdot y=x_{+}y_{-}+x_{-}y_{+}-x_{\bot}\bar{y}_{\bot}-\bar{x}_{\bot}y_{\bot}.\end{equation}
Similarly, we define derivatives in light-cone coordinates
\begin{equation}
\partial_{\pm}=\frac{\partial}{\partial x_{\mp}}=\frac{1}{\sqrt{2}}(\partial_{x_{0}}-\partial_{x_{3}}),\,\partial_{\bot}=\frac{\partial}{\partial\bar{x}_{\bot}}=\frac{1}{\sqrt{2}}(\partial_{x_{1}}+i\partial_{x_{2}}),\,\bar{\partial}_{\bot}=\frac{\partial}{\partial x_{\bot}}\end{equation}
For negative $p_{+}$ its square root is defined as $\sqrt{p_{+}}\equiv sgn(p_{+})\,\left|p_{+}\right|^{1/2}$,
therefore for this prescription $\bar{\lambda}$ picks up a minus
sign when $p$ is reversed while $\lambda$ remains unchanged:
\begin{equation}
\lambda_{\alpha}(-p)=\lambda_{\alpha}(p),\,\bar{\lambda}_{\dot{\alpha}}(-p)=-\bar{\lambda}_{\dot{\alpha}}(p).\end{equation}
To make the  light-cone notation compatible with the helicity formulation we chose 
 holomorphic and anti-holomorphic spinors as follows
\be
\lambda_\alpha =   {2^{1/4}\over \sqrt{p_+}} \left( \begin{array}{c} -{ p_{\bot}} \\
 \\
p_+ \\ \end{array} \right)\, , \qquad \bar \lambda_{\dot \alpha } = 2^{1/4} \sqrt{p_+} \left( \begin{array}{c} - {\bar p_{\bot}\over p_+} \\
\\
 1\\ \end{array} \right)\,
\label{lambda}\ee

\be
\xi_\alpha = {1\over 2^{1/4}  \sqrt{ p_+} } \left( \begin{array}{c} \sqrt  {p^2} \\
\\
 0\\ \end{array} \right)\, , \qquad \bar \xi_{\dot \alpha } =  { \sqrt{p_+}\over 2^{1/4}\, p_+} \left( \begin{array}{c}  \sqrt  {p^2}  \\
 \\
0 \\ \end{array} \right)\,.
\label{mu}\ee
Thus
\be
\lambda_\alpha \bar \lambda_{\dot \alpha }  = \sqrt 2  \left( \begin{array}{cc} {p_\bot \bar p_\bot  \over p_+}  &  -p_\bot\\
\\
-\bar p_\bot & p_+\\ \end{array} \right)\, ,
\qquad
\xi_\alpha \bar \xi_{\dot \alpha }  = \sqrt 2  \left( \begin{array}{cc} {p^2 \over 2 p_+}  &  0\\
\\
 0 & 0\\ \end{array} \right)\,
\label{mumu}\ee
and
\be
\xi_\alpha \bar \xi_{\dot \alpha }+ \lambda_\alpha \bar \lambda_{\dot \alpha }  = \sqrt 2  \left( \begin{array}{cc}  p_- &  -p_\bot \\
\\
-\bar p_\bot  & p_+\\ \end{array} \right)= p_{\alpha \dot \alpha}
 \ee
Now we may introduce the angular and square spinorial brackets
\begin{equation}
\langle p\, q\rangle\equiv\epsilon^{\alpha\beta}\lambda_{\alpha}\lambda_{\beta}=\frac{ \sqrt 2\left(p\, q\right)}{\sqrt{p_{+}}\,\sqrt{q_{+}}},\,\left[p\, q\right]\equiv\epsilon^{\dot{\alpha}\dot{\beta}}\bar{\lambda}_{\dot{\alpha}}\bar{\lambda}_{\dot{\beta}}= \sqrt 2 \left\{ p\, q\right\} \frac{\sqrt{p_{+}}\,\sqrt{q_{+}}}{p_{+}q_{+}}\label{eq:spinors}\end{equation}
where the round and curly brackets stand for
\begin{equation}
\left(p\, q\right)=p_{+}q_{\bot}-q_{+}p_{\bot},\, \qquad \left\{ p\, q\right\} =p_{+}\bar{q}_{\bot}-q_{+}\bar{p}_{\bot}.\end{equation}
Note that in terms of spinor brackets the scalar product is given
by $ 2 \, p\cdot q=\left\langle p\, q\right\rangle \left[p\, q\right]$,
which retains the correct negative sign assignment when one or both
of the momenta  flip the sign.

\section{From the light-cone superfield  action to a covariant one}
The light-cone action  \cite{Brink:1982pd}  in the real superspace basis has terms which are quadratic, cubic and quartic in superfields, see eqs. (\ref{action})-(\ref{S4}).
The anti-chiral superfield is  related to the chiral one as shown in eq. (\ref{constraint}).
We have  defined in eqs.  (\ref{eq:phi}) and (\ref{eq:phibar})
the  Fourier transforms of the light-cone superfields consistent with the constraint
({\ref{constraint}). Here we will prove that our definition of the transform in eqs.  (\ref{eq:phi}) and (\ref{eq:phibar}) is consistent with the constraint (\ref{constraint}).

With the unconstrained superfield depending on  new super-space $\phi(p,\eta)$
defined by the generalized Fourier transform (\ref{eq:phi}) of chiral
superfield, it is straightforward to derive the corresponding transformation
for anti-chiral superfield.
We substitute the expression (\ref{eq:phi}) for $\Phi(x,\theta,\bar{\theta})$
and rearrange factors associated with different SUSY indices so that
\begin{equation}
\bar{\Phi}(x,\theta,\bar{\theta})=\int\frac{d^{4}p}{(2\pi)^{4}}d^{4}\eta\, e^{ip\cdot x}\left(\frac{-i}{p_{+}^{3}}\right)\prod_{A=1}^{4}T_{A}(T_{A}^{-1}D_{A}\, T_{A})\phi(p,\eta)\end{equation}
where $T_{A}=e^{\frac{i}{2}\bar{\theta}_{A}\theta^{A}p_{+}+\eta_{A}\frac{p_{+}}{\sqrt{p_{+}}}\theta^{A}}$
is the kernel in the transformation formula for superfield $\Phi(x,\theta)$
in chiral basis. Moving SUSY covariant derivative to the right produces
a delta function
\begin{equation}
(T_{A}^{-1}D_{A}\, T_{A})\phi(p,\eta)=\left(-i\frac{p_{+}}{\sqrt{p_{+}}}\right)\delta(\bar{\theta}_{A}\sqrt{p_{+}}-i\eta_{A})\phi(p,\eta)\end{equation}
The presence of a delta function allows the Grassmann variable $\eta_{A}$
in the kernel $T_{A}$ to be replaced by $-i\bar{\theta}_{A}\sqrt{p_{+}}$
and therefore yields an overall exponent $e^{\frac{-i}{2}\bar{\theta}\cdot\theta p_{+}}$,
which is expected for an anti-chiral superfield.
\begin{equation}
\bar{\Phi}(x,\theta,\bar{\theta})=e^{\frac{-1}{2}\bar{\theta}\cdot\theta\partial_{+}}\int\frac{d^{4}p}{(2\pi)^{4}}d^{4}\eta\, e^{ip\cdot x}\,\delta^{4}(\bar{\theta}\sqrt{p_{+}}-i\eta)\left(\frac{-i}{p_{+}}\right)\phi(p,\eta)\label{eq:phibar_app}\end{equation}
From the above formula it is apparent that the definition for integral
transform (\ref{eq:phi}) is equivalent to identifying $\bar{\theta}_{A}$
with $i\eta_{A}/\sqrt{p_{+}}$ in the anti-chiral superfield in anti-chiral
basis.

In the Mandelstam formalism all anti-chiral superfields are replaced
via reality condition (\ref{constraint}). The remaining chiral superfields
$\Phi(x,\theta,\bar{\theta})$ are then rewritten in chiral basis,
which allows the $\bar{\theta}$ dependence to factor out and can
be integrated over. However this approach leads to a complicated cubic
term $S_{3}$ in the action, making the 3-point MHV vertex structure
less apparent. In this section we take another approach and use the
integral transformation formulas (\ref{eq:phi}) and (\ref{eq:phibar})
to derive the 3-point vertex directly from real basis. The cubic term
$S_{3}$ reads:
\begin{eqnarray}
&&
S_{3}=\frac{-2}{3}g\int d^{4}x\, d^{4}\theta\, d^{4}\bar{\theta}\,\frac{1}{\partial_{+}}\Phi^{a}\bar{\Phi}^{b}\partial\bar{\Phi}^{c}\, f^{abc}
\\
&&
\hspace{0.5cm} =\frac{i \sqrt 2}{3}  g\, tr\int d^{4}x\, d^{4}\theta\, d^{4}\bar{\theta}\,\prod_{i=1}^{3}\left(d^{8}z_{i}\,\phi_{i}\right)\,\left(\frac{-i}{p_{1+}}\right)\left(\frac{-i}{p_{2+}}\right)\left(\frac{-i}{p_{3+}}\right)
\nonumber \\
&&
\hspace{3cm} \times\frac{\left(p_{3\bot}-p_{2\bot}\right)}{p_{1+}}\, e^{\frac{i}{2}\bar{\theta}\cdot\theta p_{1+}+\eta_{1}\frac{p_{1+}}{\sqrt{p_{1+}}}\theta}e^{\frac{-i}{2}\bar{\theta}\cdot\theta p_{2+}}e^{\frac{-i}{2}\bar{\theta}\cdot\theta p_{3+}}
\nonumber \\
&&
\hspace{4cm}
\times e^{i\sum p_{i}\cdot x}\delta^{4}(\bar{\theta}\sqrt{p_{2+}}-i\eta_{2})\,\delta^{4}(\bar{\theta}\sqrt{p_{3+}}-i\eta_{3})
\label{eq:s3step1}\end{eqnarray}
Note that in the second line of the equation we make use the fact
that structure constant is antisymmetric in the last two indices,
$f^{abd}=-{i\over \sqrt 2} tr(T^{a}T^{b}T^{c}-T^{a}T^{c}T^{b})$, and relabel to combine
the integral into a single trace. The Fourier kernels produce an ordinary
momentum conservation delta function. Combining the remaining exponents
and integrating over $\theta$ yields $\delta^{4}(i\bar{\theta}p_{1+}+\eta_{1}\frac{p_{1+}}{\sqrt{p_{1+}}})$.
The integral is then of the form:
\begin{eqnarray}
&&
S_{3}=-\frac{\sqrt 2}{3  }g\, tr\int\prod_{i=1}^{3}\left(dz_{i}\phi_{i}^{a}\right)d^{4}\bar{\theta}\,(2\pi)^{4}\delta^{4}(\sum_{i=1}^{3}p_{i})
\nonumber \\
&&
\hspace{3cm}
\times(p_{1+}p_{3\bot}-p_{1+}p_{2\bot})\, c_{1}c_{2}c_{3}\,\left(\sqrt{p_{1+}}\sqrt{p_{2+}}\sqrt{p_{3+}}\right)^{2}
\nonumber \\
&&
\hspace{3.5cm}
\times\delta^{4}(\bar{\theta}-i\eta_{1}/\sqrt{p_{1+}})\delta^{4}(\bar{\theta}-i\eta_{1}/\sqrt{p_{2+}})\delta^{4}(\bar{\theta}-i\eta_{1}/\sqrt{p_{3+}}),\label{eq:s3step2}
\end{eqnarray}
with the phase factor $c_{i}$ representing $sgn(p_{i+})$. The subsequent
$\bar{\theta}$ integral then only contains delta functions in the
integrand and is straightforward to carry out.
\begin{equation}
\int d^{4}\theta\,\delta^{4}(\bar{\theta}-i\eta_{1}/\sqrt{p_{1+}})\delta^{4}(\bar{\theta}-i\eta_{1}/\sqrt{p_{2+}})\delta^{4}(\bar{\theta}-i\eta_{1}/\sqrt{p_{3+}})\end{equation}
\begin{equation}
=\prod_{A=1}^{4}\frac{\eta_{1A}\eta_{2A}}{\sqrt{p_{1+}}\sqrt{p_{2+}}}+\frac{\eta_{2A}\eta_{3A}}{\sqrt{p_{2+}}\sqrt{p_{3+}}}+\frac{\eta_{3A}\eta_{1A}}{\sqrt{p_{3+}}\sqrt{p_{1+}}}
=\frac{1}{(\sqrt 2)^4(12)^{4}}\prod_{A=1}^{4}\sum_{ij}\left\langle ij\right\rangle \eta_{iA}\eta_{jA},\end{equation}
In the last line of the equation above we feed a round bracket $(12)$
into super-sums \\
$\sum_{ij}\,\eta_{iA}\eta_{jA}/\sqrt{p_{i+}}\sqrt{p_{j+}}$
of each index $A$. From momentum conservation we are free to rewrite
$(12)$ as any of the other two round brackets $(12)=(23)=(31)$.
The brackets are then translated into holomorphic spinor products
according to the definition (\ref{eq:spinors}).

We note that equation (\ref{eq:s3step2}) is manifestly cyclically
symmetric, except for the factor $(p_{1+}p_{3\bot}-p_{1+}p_{2\bot})=(32)-p_{3+}p_{3\bot}+p_{2+}p_{2\bot}$
contained in the integrand. Summing over permutations eliminates the
last two terms, therefore we have:
\begin{eqnarray}
&&
S_{3}=-\frac{1}{3  }g\, tr\int\prod_{i=1}^{3}\left(dz_{i}\phi_{i}^{a}\right)d^{4}\bar{\theta}\,(2\pi)^{4}\delta^{4}(\sum_{i=1}^{3}p_{i})
\nonumber \\
&&
\hspace{3cm} \times
c_{1}c_{2}c_{3}\,\left(\sqrt{p_{1+}}\sqrt{p_{2+}}\sqrt{p_{3+}}\right)^{2}\frac{ \sqrt 2 \, (32)}{(\sqrt 2)^4(12)^{4}}\prod_{A=1}^{4}\sum_{ij}\left\langle ij\right\rangle \eta_{iA}\eta_{jA}.\end{eqnarray}
Applying the identity $(12)=(23)=(31)$ and the definition (\ref{eq:spinors})
again reproduces the familiar 3-point MHV super-vertex formula 
\begin{equation}
S_{3}=\frac{1}{3} g\, tr\int\prod_{i=1}^{3}\left(dz_{i}\phi_{i}\right)(2\pi)^{4}\delta^{4}(\sum_{i=1}^{3}p_{i})c_{1}c_{2}c_{3}\,\frac{\delta^{8}(\sum_{i}\lambda_{i}\eta_{i})}{\left\langle 12\right\rangle \left\langle 23\right\rangle \left\langle 31\right\rangle },\end{equation}
where $\delta^{8}(\sum_{i}\lambda_{i}\eta_{i})=\prod_{A=1}^{4}\sum_{ij}\left\langle ij\right\rangle \eta_{iA}\eta_{jA}$
is the Grassmannian delta function required by SUSY Ward identity.

Now we compute the $\bar{S}_{3}$ part of the SYM action
in real basis. As in the $S_{3}$ case we begin with substituting
superfields using their integral transformation formulas (\ref{eq:phi})
and (\ref{eq:phibar}). Rewriting structure constant as trace of $SU(N_{C})$
generators gives \begin{eqnarray}
&&
\bar{S}_{3}=-\frac{2}{3}g\, f^{abc}\int d^{4}x\, d^{4}\theta\, d^{4}\bar{\theta}\,\frac{1}{\partial_{+}}\bar{\Phi}^{a}\Phi^{b}\bar{\partial}\Phi
\\
&&
=\frac{i \sqrt 2}{3}g\, tr\int d^{4}x\, d^{4}\theta\, d^{4}\bar{\theta}\,\prod_{i=1}^{3}\left(d^{8}z_{i}\,\phi_{i}\right)\,\left(\frac{-i}{p_{1+}}\right)\left(\frac{-i}{p_{2+}}\right)\left(\frac{-i}{p_{3+}}\right)
e^{i\sum p_{i}\cdot x}\frac{\left(\bar{p}_{3\bot}-\bar{p}_{2\bot}\right)}{p_{1+}}
\nonumber \\
&&
\hspace{3cm} \times e^{\frac{i}{2}\bar{\theta}\cdot\theta p_{2+}+\eta_{2}\frac{p_{2+}}{\sqrt{p_{2+}}}\theta}e^{\frac{i}{2}\bar{\theta}\cdot\theta p_{3+}+\eta_{1}\frac{p_{3+}}{\sqrt{p_{3+}}}\theta}e^{\frac{-i}{2}\bar{\theta}\cdot\theta p_{1+}}\delta^{4}(\bar{\theta}\sqrt{p_{1+}}-i\eta_{1}).\label{eq:s3bstep1}
\end{eqnarray}
We then integrate over spacetime coordinates to produce the momentum
conservation delta function, which along with the Grassmannian delta
function $\delta^{4}(\bar{\theta}\sqrt{p_{1+}}-i\eta_{1})$ put the
last line of equation (\ref{eq:s3bstep1}) into the form
\begin{equation}
e^{(\eta_{1}\frac{p_{1+}}{\sqrt{p_{1+}}}+\eta_{2}\frac{p_{2+}}{\sqrt{p_{2+}}}+\eta_{3}\frac{p_{3+}}{\sqrt{p_{3+}}})\theta}\delta^{4}(\bar{\theta}\sqrt{p_{1+}}-i\eta_{1}).\end{equation}
Integrating over $\theta$ and $\bar{\theta}$ yields
\begin{eqnarray}
&&
\bar{S}_{3}=\frac{i\sqrt 2}{3}g\, tr\int\prod_{i=1}^{3}\left(dz_{i}\phi_{i}\right)\,\left(\frac{-i}{p_{1+}}\right)\left(\frac{-i}{p_{2+}}\right)\left(\frac{-i}{p_{3+}}\right)
\nonumber \\
&&
\hspace{1.5cm} \times(2\pi)^{4}\delta^{4}(\sum_{i=1}^{3}p_{i})\,\left(p_{1+}\bar{p}_{3\bot}-p_{1+}\bar{p}_{2\bot}\right)\,\delta^{4}(\eta_{1}\frac{p_{1+}}{\sqrt{p_{1+}}}+\eta_{2}\frac{p_{2+}}{\sqrt{p_{2+}}}+\eta_{3}\frac{p_{3+}}{\sqrt{p_{3+}}}).\label{eq:s3bstep2}
\end{eqnarray}
Again we cyclically symmetrize the integrand, this leaves only the
curly bracket in $\left(p_{1+}\bar{p}_{3\bot}-p_{1+}\bar{p}_{2\bot}\right)=\left\{ 32\right\} -p_{\text{3+}}\bar{p}_{3\bot}+p_{2+}\bar{p}_{2\bot}$.
To put the above expression (\ref{eq:s3bstep2}) into more familiar
form, we rewrite the delta function $\delta^{4}(\eta_{1}\frac{p_{1+}}{\sqrt{p_{1+}}}+\eta_{2}\frac{p_{2+}}{\sqrt{p_{2+}}}+\eta_{3}\frac{p_{3+}}{\sqrt{p_{3+}}})$
as
\begin{eqnarray}
&&
\left(\frac{p_{1+}p_{2+}p_{3+}}{\sqrt{p_{1+}}\sqrt{p_{2+}}\sqrt{p_{3+}}}\right)^{4}\delta^{4}(\eta_{1}\frac{\sqrt{p_{2+}}\sqrt{p_{3+}}}{p_{2+}p_{3+}}+\eta_{2}\frac{\sqrt{p_{3+}}\sqrt{p_{1+}}}{p_{3+}p_{1+}}+\eta_{3}\frac{\sqrt{p_{1+}}\sqrt{p_{2+}}}{p_{1+}p_{2+}})
\\
&&
=\left(\frac{p_{1+}p_{2+}p_{3+}}{\sqrt{p_{1+}}\sqrt{p_{2+}}\sqrt{p_{3+}}}\right)^{4}\frac{1}{ (\sqrt 2)^4\left\{ 12\right\} ^{4}}\delta^{4}(\eta_{1}\left[23\right]+\eta_{2}\left[31\right]+\eta_{3}\left[12\right]),\end{eqnarray}
where we take the definition of anti-holomorphic spinor product as
$\left[p\, q\right]= \sqrt 2 \left\{ p\, q\right\} \sqrt{p_{+}}\sqrt{q_{+}}/p_{+}q_{+}$
and we use the bilinear property of curly bracket $ \left\{ 12\right\} =\left\{ 23\right\} =\left\{ 31\right\} $.
After cancellation with one curly bracket coming from the factor $\left(p_{1+}\bar{p}_{3\bot}-p_{1+}\bar{p}_{2\bot}\right)$
we rearrange the remaining ones in the denominator as a sequential product
$\left\{ 12\right\} \left\{ 23\right\} \left\{ 31\right\} $. Translating
curly brackets into spinor products then gives
\begin{equation}
\bar{S}_{3}=\frac{1}{3}g\, tr\int\prod_{i=1}^{3}\left(dz_{i}\phi_{i}\right)\delta^{4}(\sum_{i=1}^{3}p_{i})c_{1}c_{2}c_{3}\,\frac{\delta^{4}(\eta_{1}\left[23\right]+\eta_{2}\left[31\right]+\eta_{3}\left[12\right])}{\left[12\right]\left[23\right]\left[31\right]}.\end{equation}



\section {A  quartic light-cone superfield action  in new variables}

The 4-point interaction in the action is split into two parts according
to whether the anti-chiral superfields are adjacent:
First we compute the adjacent part using integral transformations
eqs.  (\ref{eq:phi}) and (\ref{eq:phibar})
\begin{eqnarray}
&&
S_{4}^{1}=\frac{1}{4}g^{2}tr\int d^{4}\theta\, d^{4}\bar{\theta}\left(\prod_{i=1}^{4}d^{8}z_{i}\phi_{i}\right)(2\pi)^{4}\delta(\sum_{i}p_{i})
\nonumber \\
&&
\times\left(\frac{-i}{p_{1+}}\right)\left(\frac{-i}{p_{2+}}\right)\left(\frac{-i}{p_{3+}}\right)\left(\frac{-i}{p_{4+}}\right)\frac{(p_{1+}-p_{2+})}{(p_{1+}+p_{2+})}\,\frac{(p_{3+}-p_{4+})}{(p_{3+}+p_{4+})}
\nonumber \\
&&
\times e^{\frac{i}{2}\bar{\theta}\cdot\theta p_{1+}+\eta_{1}\frac{p_{1+}}{\sqrt{p_{1+}}}\theta}e^{\frac{i}{2}\bar{\theta}\cdot\theta p_{2+}+\eta_{2}\frac{p_{2+}}{\sqrt{p_{2+}}}\theta}
\nonumber \\
&&
\times e^{-\frac{i}{2}\bar{\theta}\cdot\theta p_{3+}}\delta^{4}(\bar{\theta}\sqrt{p_{3+}}-i\eta_{3})e^{-\frac{i}{2}\bar{\theta}\cdot\theta p_{4+}}\delta^{4}(\bar{\theta}\sqrt{p_{4+}}-i\eta_{4})
\label{eq:s4step1}
\end{eqnarray}

In order to simplify the super-vertex formula in the $\theta$ $\bar{\theta}$
integral we use momentum conservation to replace $\frac{i}{2}\bar{\theta}\cdot\theta p_{1+}+\frac{i}{2}\bar{\theta}\cdot\theta p_{2+}$
in the exponents by $-\frac{i}{2}\bar{\theta}\cdot\theta p_{3+}-\frac{i}{2}\bar{\theta}\cdot\theta p_{4+}$.
The last two lines of the equation (\ref{eq:s4step1}) give
\begin{eqnarray}
&&
\int d^{4}\theta\, d^{4}\bar{\theta}\, e^{-i\bar{\theta}\cdot\theta p_{3+}}e^{-i\bar{\theta}\cdot\theta p_{4+}}e^{\eta_{1}\frac{p_{1+}}{\sqrt{p_{1+}}}\theta}e^{\eta_{2}\frac{p_{2+}}{\sqrt{p_{2+}}}\theta}
\nonumber \\
&&
\hspace{2cm} \times\delta^{4}(\bar{\theta}\sqrt{p_{3+}}-i\eta_{3})\delta^{4}(\bar{\theta}\sqrt{p_{4+}}-i\eta_{4})
\\
&&
=\int d^{4}\theta\, d^{4}\bar{\theta}\, e^{(\sum_{i}\,\eta_{i}\frac{p_{i+}}{\sqrt{p_{i+}}})\theta}\delta^{4}(\bar{\theta}\sqrt{p_{3+}}-i\eta_{3})\delta^{4}(\bar{\theta}\sqrt{p_{4+}}-i\eta_{4})
\\
&&
=\delta^{4}(\sum_{i}\eta_{i}\frac{p_{i+}}{\sqrt{p_{i+}}})\,\delta^{4}(\eta_{3}\sqrt{p_{4+}}-\eta_{4}\sqrt{p_{3+}})
\label{eq:pibracket}
\end{eqnarray}
We obtain the vertex by cyclically symmetrizing the above result
\begin{equation}
S_{4}^{1}=tr\int\left(\prod_{i=1}^{4}d^{8}z_{i}\phi_{i}\right)(2\pi)^{4}\delta(\sum_{i}p_{i})\delta^{4}(\sum_{i}\eta_{i}\frac{p_{i+}}{\sqrt{p_{i+}}})\, V_{4}^{1},\end{equation}
with 
\begin{eqnarray}
&&
V_{4}^{1}=-\frac{g^2}{8}\,\frac{1}{p_{1+}p_{2+}p_{3+}p_{4+}}\,\frac{(p_{1+}-p_{2+})}{(p_{1+}+p_{2+})}\,\frac{(p_{3+}-p_{4+})}{(p_{3+}+p_{4+})}
\nonumber \\
&&
\hspace{2cm} \times \left(\delta^{4}(\eta_{1}\sqrt{p_{2+}}-\eta_{2}\sqrt{p_{1+}})+\delta^{4}(\eta_{3}\sqrt{p_{4+}}-\eta_{4}\sqrt{p_{3+}})\right)
\nonumber \\
&&
\hspace{1cm} + cycl.\, perm.
\end{eqnarray}
Since the $\theta$ $\bar{\theta}$ integral in $S_{4}^{2}$ is the
same as (\ref{eq:pibracket}) provided one relabels $2\rightarrow3$
and $3\rightarrow2$ in the expression, it is easy to see that
\begin{eqnarray}
&&
V_{4}^{2}=-\frac{g^{2}}{16}\,\frac{1}{p_{1+}p_{2+}p_{3+}p_{4+}}\left(\delta^{4}(\eta_{2}\sqrt{p_{4+}}-\eta_{4}\sqrt{p_{2+}})+\delta^{4}(\eta_{1}\sqrt{p_{3+}}-\eta_{3}\sqrt{p_{1+}})\right.
\nonumber \\
&&
\hspace{0.5cm} \left.-\delta^{4}(\eta_{1}\sqrt{p_{4+}}-\eta_{4}\sqrt{p_{1+}})-\delta^{4}(\eta_{2}\sqrt{p_{3+}}-\eta_{3}\sqrt{p_{2+}})+cycl.\, perm. \right)
\end{eqnarray}


\end{document}